\PassOptionsToPackage{table,xcdraw,dvipsnames}{xcolor}
\documentclass[
]{revtex4-2}
\usepackage{adjustbox}
\usepackage{changepage}
\usepackage{amsmath}
\usepackage{float}
\usepackage{tikz}
\usepackage{wrapfig}
\usepackage{graphicx}
\usepackage{dcolumn}
\usepackage[linesnumbered,ruled,vlined]{algorithm2e}
\usepackage{algpseudocode}
\usepackage{epstopdf}
\usepackage{appendix}
\usepackage{cases}
\usepackage{setspace}
\usepackage{lineno}
\usepackage{bm}

\newcommand{\beginsupplement}{%
    \setcounter{table}{0}
    \renewcommand{\thetable}{S\arabic{table}}%
    \setcounter{figure}{0}
    \renewcommand{\thefigure}{S\arabic{figure}}%
    \setcounter{equation}{0}
    \renewcommand{\theequation}{S\arabic{equation}}%
}
\begin{document}

\preprint{APS/123-QED}

\title{Transient segregation of bi-disperse granular mixtures in a periodic chute flow 
}

\author{Soniya Kumawat}
\affiliation{
 Department of Chemical Engineering, Indian Institute of Technology Kanpur, Uttar Pradesh, 208016, India
}
\author{Vishnu Kumar Sahu}
\affiliation{
 Department of Chemical Engineering, Indian Institute of Technology Kanpur, Uttar Pradesh, 208016, India
}
\author{Anurag Tripathi}%
\email{anuragt@iitk.ac.in}
\affiliation{
 Department of Chemical Engineering, Indian Institute of Technology Kanpur, Uttar Pradesh, 208016, India
}

\date{\today}

\begin{abstract}
Transient size segregation of a bi-disperse granular mixture flowing over a periodic chute is studied using the Discrete Element Method and continuum simulations. A recently developed particle force-based size segregation model is used to predict the time-dependent flow properties of binary mixtures starting from rest.  
A two-way coupled continuum model that solves the momentum balance and convection-diffusion equations by incorporating the mixture segregation model along with the generalized inertial number-based rheological model is developed for predicting the evolution of segregation.
The predicted concentration profiles and other flow properties of the mixture are found to be in good agreement with the DEM data for a variety of compositions. The evolution of the centre of mass of the two species with time is also very well captured for different initial configurations and size ratios using the particle force-based segregation model.

\end{abstract}

\maketitle

Granular mixtures consisting of particles differing in size undergo segregation during their flow. Understanding the evolution of segregation is important for predicting granular material behavior in geophysical situations and in optimizing industrial processes involving particulate mixtures. The segregation of granular mixtures has been extensively investigated in the past few decades and detailed review about granular segregation has been summarized in \cite{gray2018particle,umbanhowar2019modeling}. 
The popular methods for proposing segregation models are based on kinetic theory~\cite{jenkins1985kinetic,savage1988particle,larcher2013segregation,larcher2015evolution,jenkins2020segregation}, empirical correlations~\cite{fan2014modelling,schlick2015granular,jones2018asymmetric,Duan2021}, and particle force-based approaches~\cite{jing2020rising,tripathi2021size,yennemadi2023drag}. 
Steady-state segregation in different geometries have been predicted using each of these approaches~\cite{larcher2013segregation,tripathi2013density,fan2014modelling,schlick2015granular,xiao2016modelling,Duan2021,tripathi2021size}.
The theoretical aspects of time-dependent size segregation in free surface flow was first qualitatively explored by ~\cite{grayandchugunov2006particle,gray_ancey_2011}, which was utilised by \cite{marks_rognon_einav_2012} to predict the time evolution of segregation for polydisperse size mixtures flowing over a periodic chute. 
Later studies~\cite{fan2014modelling,schlick2016} refined the segregation model to account for the dependency on the local shear rate and predicted the time-dependent concentration of species in polydisperse mixtures flowing over bounded heap and rotating tumblers~\cite{deng2018continuum}. 

Barker et. al.~\cite{barker2021OpenFoam} developed a $2D$ continuum model by using the segregation model by \cite{trewhela2021experimental} that accounts for the shear rate as well as pressure dependence of the segregation velocity along with the regularized $\mu(I)$ rheological model of \cite{barker_gray_2017}. The study was able to predict the qualitative behavior of size segregation for two different flow geometries, inclined chute and rotating box. The same continuum model was also utilized for predicting the size segregation in triangular drum~\cite{maguire2024particle}, without any comparison with DEM or experimental data. A one-dimensional continuum model was developed by~\cite{trewhela2024segregation} to predict the time evolution of size segregation in plane shear flow by integrating the segregation model from \cite{trewhela2021experimental} with the regularized $\mu(I)$ rheological model of \cite{barker_gray_2017} to obtain qualitative predictions. 
Furthermore, \cite{liuandhennan2023coupled} developed a phenomenological continuum model to predict size segregation driven by strain-rate gradients for dense bi-disperse granular mixtures of disks and spheres in vertical chute flow and annular shear flow. In a later study, \cite{singhandhennan2024continuum} extended this phenomenological model to account for the presence of pressure-gradient-driven segregation as well. The authors incorporated non-local rheological model to enable describing quasi-static flow regime using their continuum framework. The continuum model predictions at different times were found to be in good agreement with the DEM data. 
While kinetic theory-based formulations have also been used to explore the time evolution of segregation driven by size differences~\cite{larcher2015evolution}, these segregation models provide reasonable predictions only for very small size differences. 

Recently, particle force-based segregation model has been used to predict evolution of segregation in different density granular mixtures in a periodic chute~\cite{Kumawat_Sahu_Tripathi_2025}. Following a similar approach, we present a $1D$ continuum model for transient size segregation in bi-disperse granular mixtures using particle force-based size segregation model~\cite{tripathi2021size,yennemadi2023drag}.
Our continuum model predicts transient species concentration profiles by accounting for the evolving flow kinematics with time, which requires solving the momentum balance equations coupled with the granular mixture rheology. 
We consider a variety of mixture compositions as well as initial configurations of the two species and show that the model is accurately able to predict segregation evolution in various situations. 

We consider a binary granular mixture of large and small species having concentrations $f_L$ and $f_s$, respectively, flowing over an inclined surface. Following previous studies~\cite{savage1988particle,gray_ancey_2011,fan2014modelling,xiao2016modelling,deng2019modeling}, we describe the evolution of the concentration of the $i^{th}$ species ($f_i$) using the advection-diffusion-segregation equation,
\begin{equation}
   \frac{\partial f_i}{\partial t} +   \nabla \cdot (\textbf{v} f_i) +  \nabla \cdot (\textbf{J}_{i}^{S}) + \nabla \cdot (\textbf{J}_{i}^{D}) = 0,
   \label{eq:adv_diff_seg_vectoreqn}
\end{equation}
where, $\textbf{v}$ is the mixture velocity and $\textbf{J}_{i}^{S}$ and $\textbf {J}_{i}^{D}$ are segregation and diffusion fluxes of $i^{th}$ species, respectively. For the periodic system used in our DEM simulations, only the gradient along the vertical ($y$) direction is non-zero, and we consider unidirectional flow over an inclined plane. With these assumptions, segregation and diffusion fluxes along the $y$ direction are defined as $J_{i}^{S} = v_i^{seg} f_i$ and $J_i^{D} = -D\frac{\partial f_i} {\partial y}$, and the convective term becomes zero.
Substituting the segregation flux expressions into equation~\ref{eq:adv_diff_seg_vectoreqn} leads to
\begin{equation}
   \frac{\partial f_i}{\partial t} + \frac{\partial }{\partial y } ( v_i^{seg} f_i) =  \frac{\partial}{\partial y} \left(D\frac{\partial f_i} {\partial y}\right).
   \label{eq:Conc_pde}
\end{equation}
In line with previous studies~\cite{tripathi2021size,sahu_kumawat_agrawal_tripathi_2023}, we utilize the linear variation of the diffusivity ($D$) with shear rate ($\dot \gamma $) and volume average diameter ($d_{mix}$) of the mixture, i.e., $D = b\dot \gamma d_{mix}^2$, where $b$ is a constant.
The segregation velocity ($v_i^{seg}$) of a large size species in a binary mixture is obtained using the particle force-based approach proposed by \cite{tripathi2021size}
\begin{equation}
    v_L^{seg} = \frac{m_L g cos\theta}{c \pi \eta d_L} \alpha (1-f_L)(1 + k f_L)
\end{equation}
where $m_L$ is the mass of the particles of large species in the mixture and $\eta = |\tau_{yx}|/\dot \gamma$ is the effective mixture viscosity.
The Stokes' drag coefficient $c$ for different solids factions ($\phi$) is taken from the simulation study by \cite{tripathi2013density}. Here, $\alpha$ represents the ratio of the total upward force to the weight of the large particle. 
Yennemadi and Khakhar(\cite{yennemadi2023drag}) have shown that it can be expressed as $\alpha = [F(r) - 1][1 + (A - B r) \tan\theta]$, where the term $(A - B r) \tan\theta$ accounts for the contribution of the lift force. \textcolor{black}{Since the values of parameters $A$ and $B$ are $0.027$ and $0.028$, respectively, the contribution of lift force to the total upward force is less than $10\%$.}
The function $F(r)$ for single large intruder of size ratio $r$ is given by \cite{jing2020rising} as $F(r) = \phi [1 - c_1 \exp{(-r/R1)}] [1 + c_2 \exp{(-r/R2)}]$,
where $\phi$ is the local solids fraction. By carefully calculating the total upward force on a large size intruder, \cite{yennemadi2023drag} refitted this functional form and obtained the fitting coefficients as $c_1 = 20.3$, $c_2 = 2.11$, $R_1 = 0.25$, and $R_2 = 4.17$. We utilize this functional form to calculate $\alpha$ and note that the values of $\alpha$ for a single intruder obtained by \cite{tripathi2021size} are close to the values reported by \cite{yennemadi2023drag}. 

By substituting the expressions for diffusivity and the segregation velocity of the large species in equation~\ref{eq:Conc_pde}, we obtain a non-linear partial differential equation in large species concentration ($f_L$). 
This equation accounts for the effects of local shear rate and viscosity on the concentration evolution. Since the viscosity of dense granular flow depends on both the local shear rate and pressure, prediction of species concentration requires knowledge of the flow kinematics as well as pressure. 
For this, we solve the momentum balance equations for unidirectional, fully developed flow over a surface inclined at an angle $\theta$, which simplify to
\begin{equation}
   \rho_b \frac{\partial v_x }{\partial  t} =   \rho_b g \sin\theta - \frac{\partial \tau_{yx}}{\partial y},
   \label{eq:mombal_x}
\end{equation}
\begin{equation}
   P(y,t) =  g \cos \theta (1 - a \tan \theta )\int_{y}^{h}  \rho_b(y,t) dy.
    \label{eq:mombal_y}
\end{equation}
Here, $v_x$ represents the mixture velocity in the $x-$ direction, $\rho_b$ is the bulk density, $\tau_{yx}$ denotes the shear stress and $y = 0$ and $y = h$ represent the chute base and free surface, respectively. For mixtures of different-sized particles with the same density, the bulk density is given by $\rho_b = \phi \rho_p$, where $\phi$ is the local solids fraction and $\rho_p$ is the particle density. The shear stress ($\tau_{yx}$) and pressure ($P$) relate through the inertial number-dependent effective friction coefficient, defined as $\mu(I_{mix}) = |\tau_{yx}|/P$. 
The effective friction coefficient and local solids fraction are obtained using the JFP model and dilatancy law~\cite{jop2006constitutivenature}, respectively. We follow the generalised inertial number approach of \cite{tripathi2011rheology} for granular mixtures and also account for the presence of the stress anisotropy using parameter ($a$). In addition, we account for the solids fraction dependency on the species concentration~\cite{tripathi2011rheology}. The values of the various parameters are reported in \ref{sec:mu_I_phi_I}.
We solve the segregation-diffusion equation (Eq~\ref{eq:Conc_pde}) and momentum balance equation (Eq~\ref{eq:mombal_x}) simultaneously using the PDEPE solver along with the appropriate initial and boundary conditions to predict the species concentration and velocity fields.  More details about the numerical method for solving these equations are given in \ref{sec:NumericalMethod}. To enable comparison of our model predictions with Discrete Element Method (DEM), we also simulate spherical particles of different sizes and the same density ($\rho_p$) flowing down an inclined plane, details of which are given in \ref{sec:simulationMethod}. \textcolor{black}{The results are reported in dimensionless form using small particle diameter $d$ as the length scale, $(d/g)^{1/2}$ as the time scale, and $(gd)^{1/2}$ as the velocity scale. All the results are reported for $\theta = 25^\circ$.}


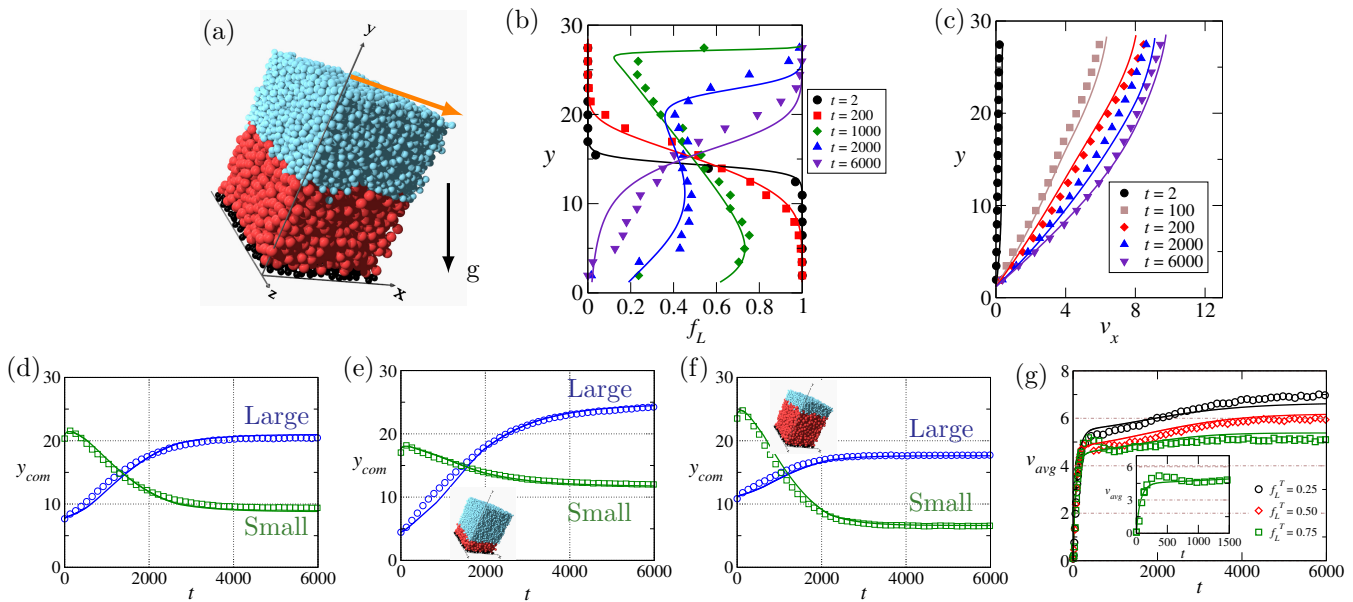
\begin{figure}[h]
    \centering
      \begin{minipage}{0.32\textwidth}
        \centering
       \adjustbox{raise=0.5cm}{  \includegraphics[scale=0.13, trim=200 0 30 0, clip]{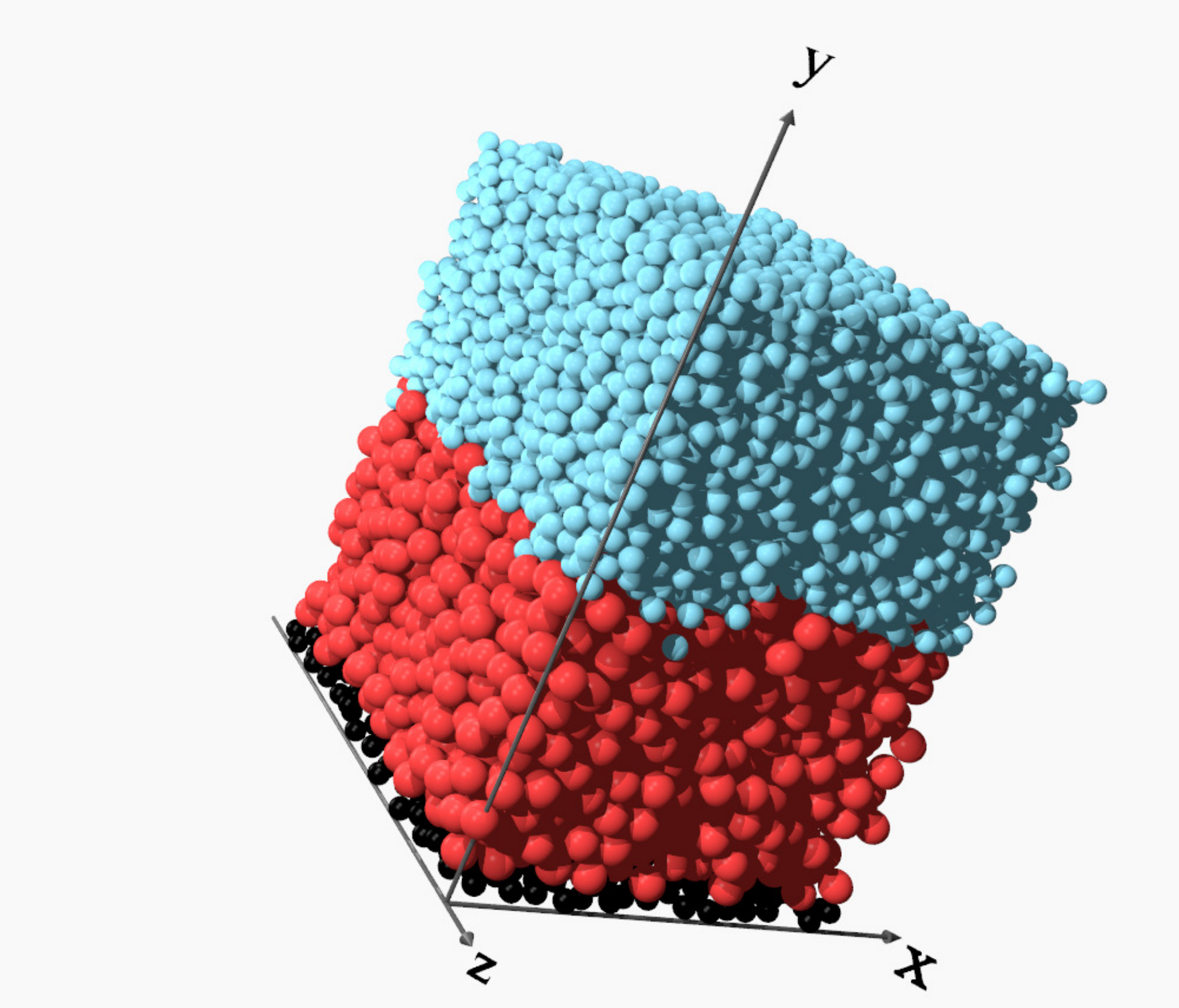}}\put(-100,115){(a)}\put(0,25){g}
      \end{minipage}
      \begin{tikzpicture}[overlay, remember picture]
         \draw[->, orange, very thick, line width=1.7pt, >=latex] (-2.7,1.42) -- (-1.2,0.9); 
          \draw[->, black, very thick, line width=1.2pt, >=latex] (-1.4,0.0) -- (-1.4,-1.2);
    \end{tikzpicture}
      \hspace{-1cm}
      \begin{minipage}{0.32\textwidth}
        \centering
    \includegraphics[scale=0.33]{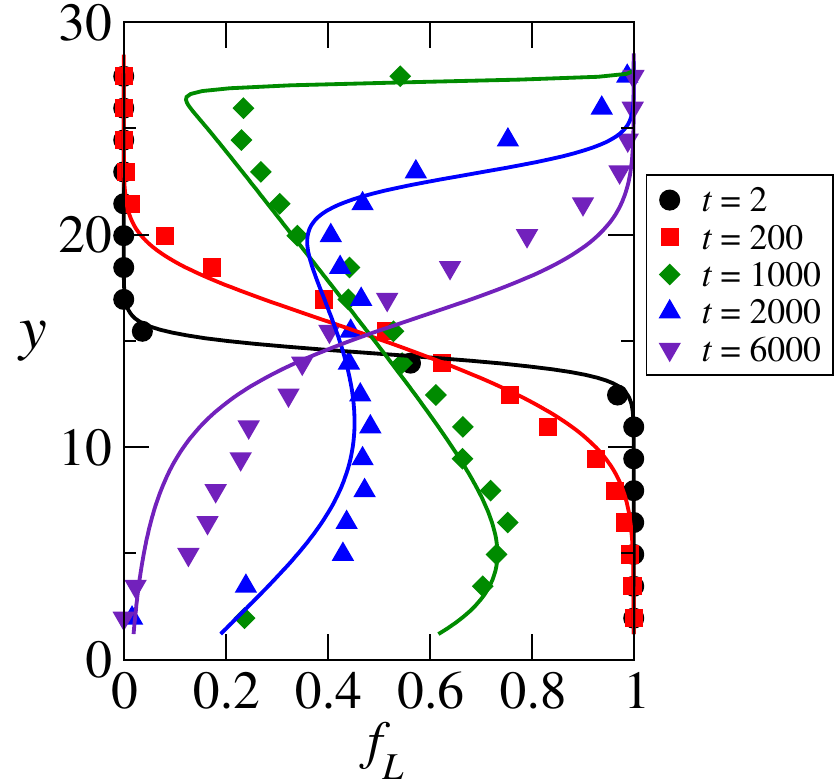}
    \put(-145,120){(b)}
     \end{minipage}
  \hspace{-1cm}
    \begin{minipage}{0.32\textwidth}
        \centering
    \includegraphics[scale=0.31]{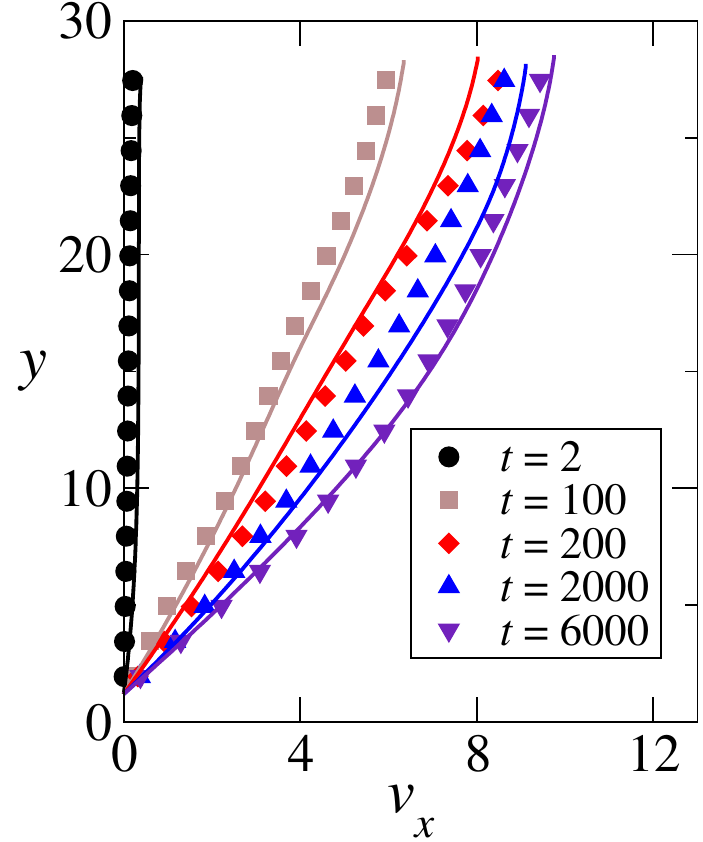}\put(-110,120){(c)}
     \end{minipage}
    \hfill
   \begin{adjustwidth}{-1cm}{-0.5cm}
   \begin{minipage}{0.32\textwidth}
        \centering
    \includegraphics[scale=0.26]{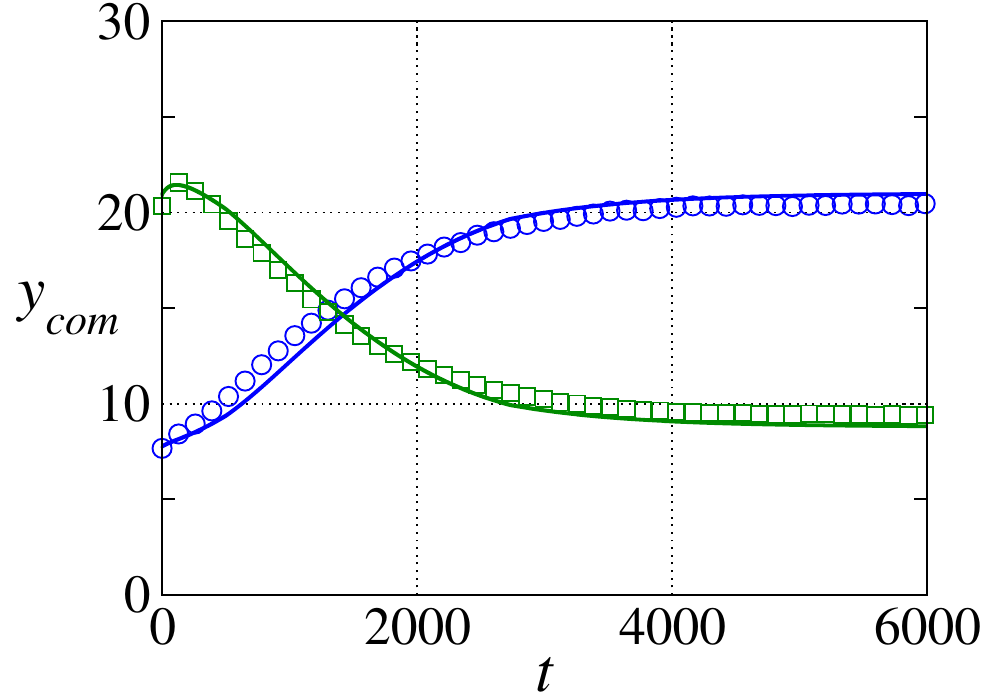}\put(-125,85){(d)}\put(-35,25){\textcolor{OliveGreen}{Small}}\put(-35,67){\textcolor{Blue}{Large}}
     \end{minipage}
      \hspace{-1.5cm}
      \begin{minipage}{0.32\textwidth}
        \centering
 \begin{tikzpicture}
        \node[anchor=south west] (base) at (0,0) {\includegraphics[scale=0.26]{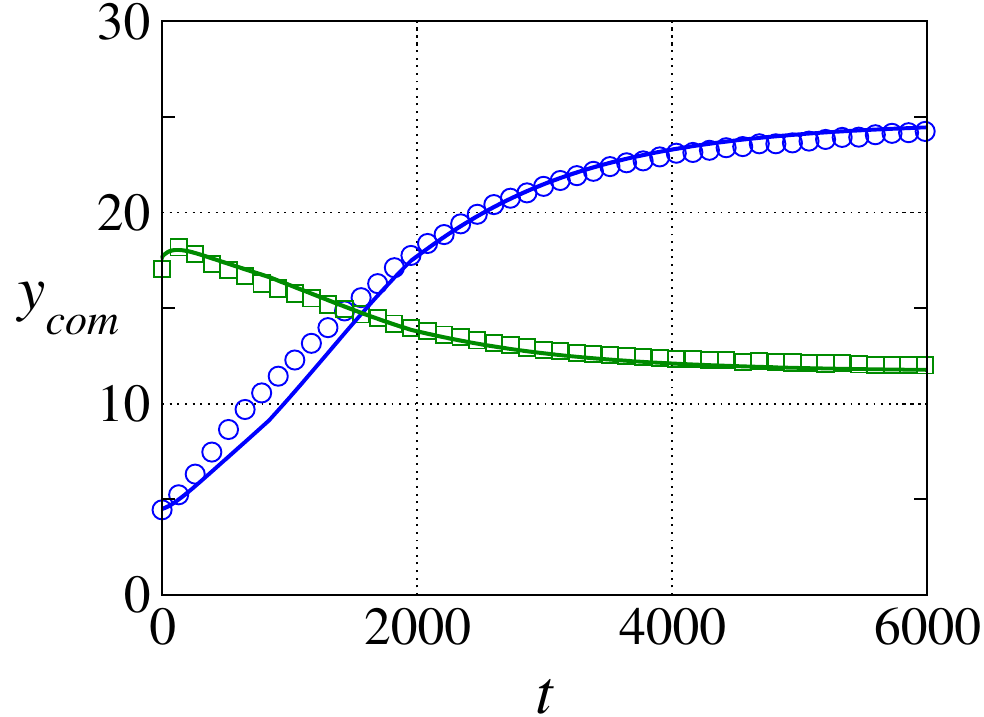}\put(-125,85){(e)}
        \put(-40,30){\textcolor{OliveGreen}{Small}}\put(-40,77){\textcolor{Blue}{Large}}
        };
        \node[anchor=south west] at (1.37,0.545) {\includegraphics[scale=0.032, trim=200 0 30 0, clip]{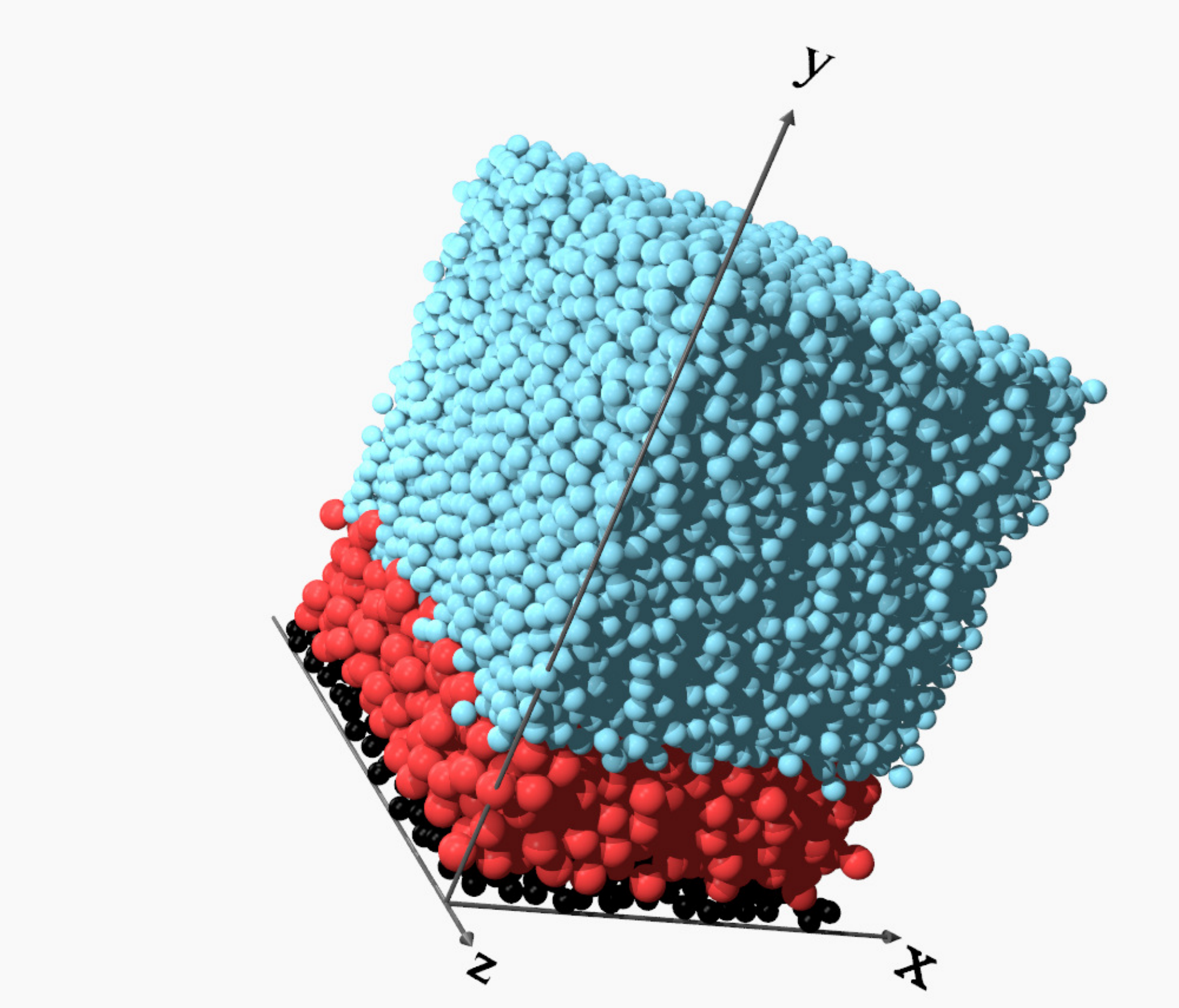}};
    \end{tikzpicture} 
    \end{minipage}
     \hspace{-1.5cm}
     \begin{minipage}{0.32\textwidth}
        \centering
 \begin{tikzpicture}
        \node[anchor=south west] (base) at (0,0) {\includegraphics[scale=0.26]{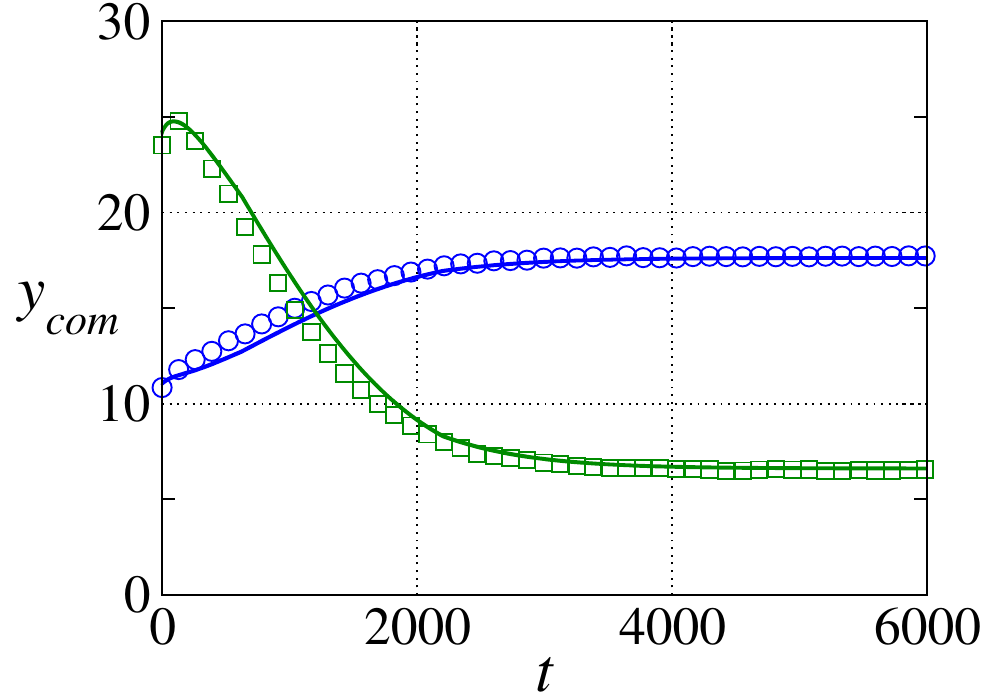}\put(-125,85){(f)}
        \put(-38,34){\textcolor{OliveGreen}{Small}}\put(-38,62){\textcolor{Blue}{Large}}
        };
        \node[anchor=south west] at (1.16,1.96) {\includegraphics[scale=0.032, trim=200 0 30 0, clip]{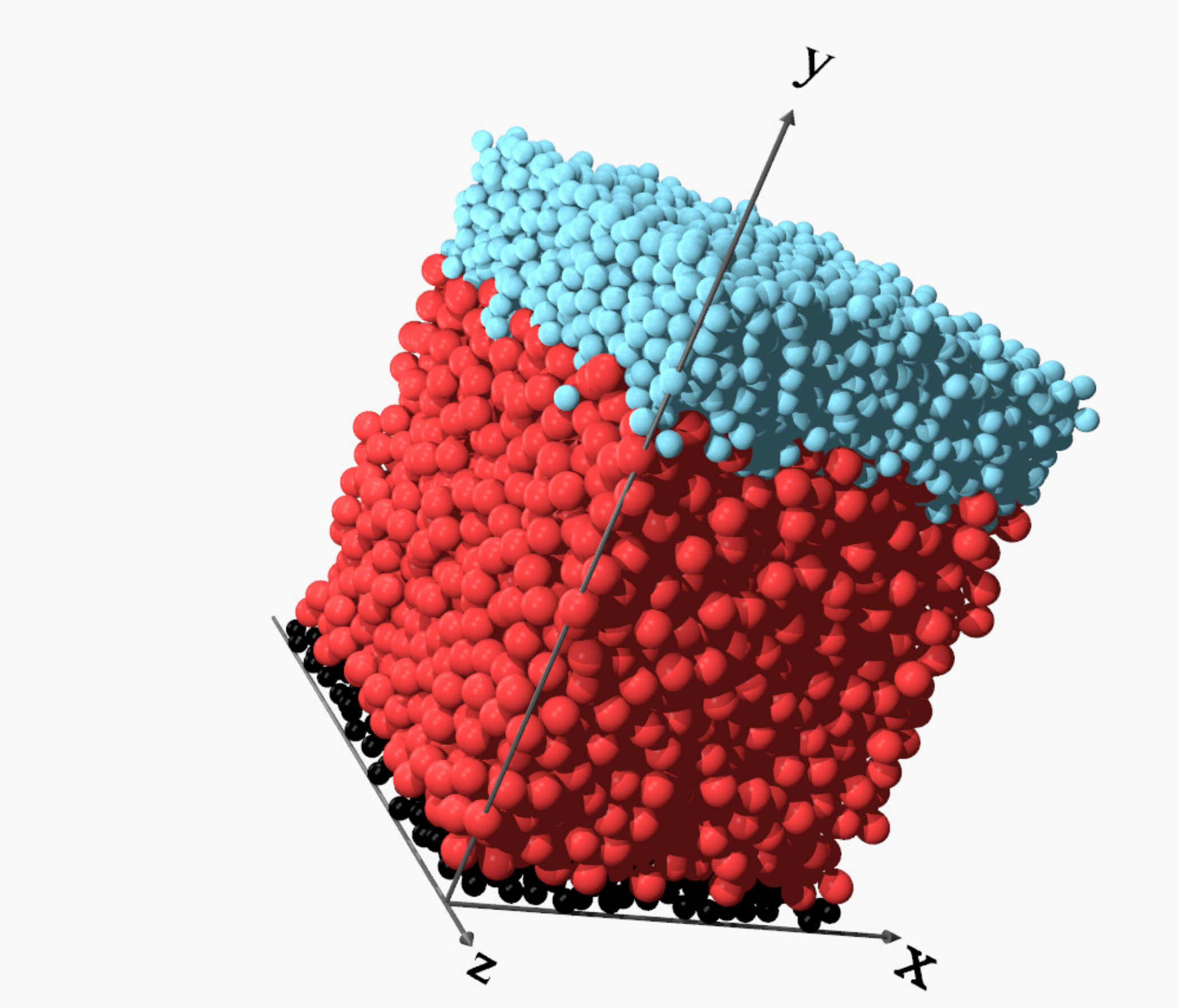}};
    \end{tikzpicture} 
      \end{minipage}
       \hspace{-1.5cm}
     \begin{minipage}{0.32\textwidth}
        \centering
     \includegraphics[scale=0.26]{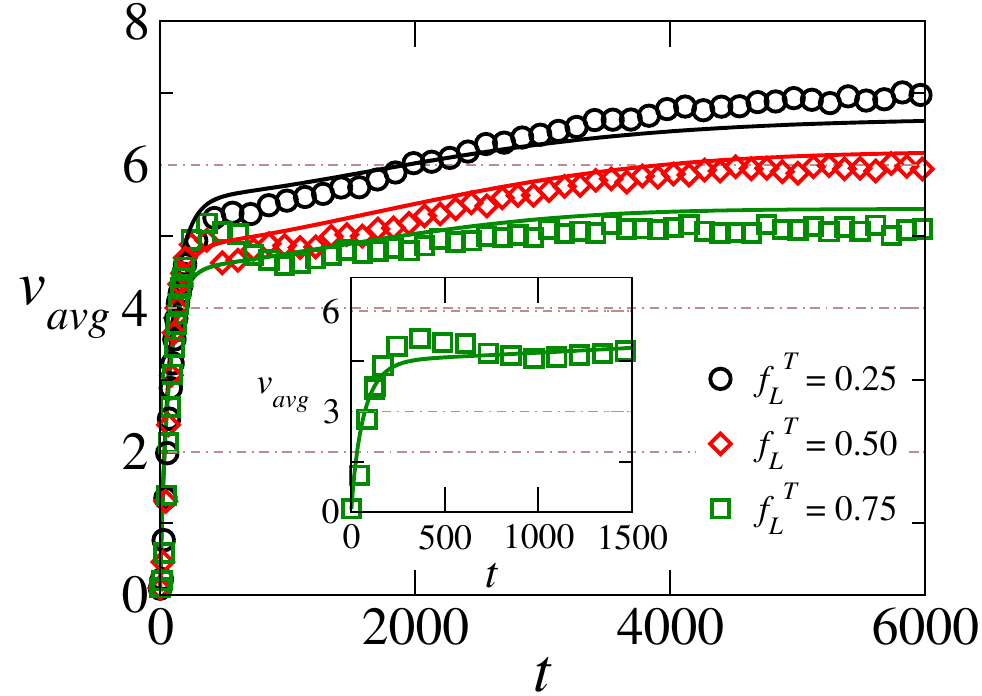}
     \put(-125,80){(g)}
      \end{minipage}
      \end{adjustwidth}
    \caption{(a) DEM snapshot for initial state starting from LNB case for binary mixture consisting of $50\%$ large (red) and $50\%$ small (blue) particles. Static (black) particles are used to form a rough bumpy base to prevent slip at the bottom of the chute. Instantaneous profiles for (b) concentration of large species, and (c) mixture velocity for binary mixture of $50\%$ large and small particles having size ratio $r = 1.5$ flowing over inclination angle $\theta = 25^o$. Variation of centre of mass of both large and small species for three different mixture compositions, (d) $f^T_L = 0.50$, (e) $f^T_L = 0.25$, and (f) $f^T_L = 0.75$. (g) Time variation of average mixture velocity for all three different composition mixtures.}  
    \label{fig:r_1.5_theta_25_instant}
\end{figure}
Figure~\ref{fig:r_1.5_theta_25_instant} shows the results for the binary mixture of size ratio $r = 1.5$ started from rest at inclination angle $\theta=  25^o$. Symbols represent the DEM data, while solid lines correspond to the continuum model predictions. The flow begins with large particles near the base (LNB) configuration  (figure~\ref{fig:r_1.5_theta_25_instant}a) for a binary mixture consisting of equal volume of large and small particles ($f^T_L = 0.5$). Figure~\ref{fig:r_1.5_theta_25_instant}b shows the variation of large species concentration $f_L$ across the layer at different times. 
The large species concentration profile at $t = 2$ (black line) exhibits a behavior resembling that of a sigmoid function.
As time progresses, large particles rise towards the free surface while small particles settle near the base, causing the concentration profile of large species to increase near surface and decrease near the base. The system reaches a steady state at time $t \approx 6000$, after which concentration profile remains unchanged. The continuum model predictions (lines) capture this evolution of large species concentrations over time and space and show good agreement with the DEM data (symbols). Figure~\ref{fig:r_1.5_theta_25_instant}c shows the variation of mixture velocity $v_x$ with distance $y$ from the base at different time instants. 
The flow starts from rest i.e., with velocity $v_x = 0$  throughout the layer. With time, the velocity profile evolves to show Bagnold-like behavior, with $v_x \to 0$ at the base and maximum velocity at the free surface. For $t > 2000$ time units, velocity changes become very gradual, while the segregation keeps evolving, indicating that the rheological time scale is smaller than the segregation time scale. The continuum model predictions for mixture velocity ($v_x$) show good agreement with the DEM data (symbols).
Predictions for other flow properties such as shear stress and inertial number also match well with DEM data (see figure~\ref{fig:r_1.5_theta_25_instant_flowProp} of the supplementary material).

Figures~\ref{fig:r_1.5_theta_25_instant}d-\ref{fig:r_1.5_theta_25_instant}f show the variation of species centre of mass position $y_{com}$ for mixtures having $f^T_L = 0.50$, $f^T_L = 0.25$, and $f^T_L = 0.75$, respectively. DEM snapshots for $f^T_L = 0.25$ and $f^T_L = 0.75$ are shown in the insets of the corresponding figures. The continuum model predictions match well with the DEM simulation data for these diverse range of compositions. The average mixture velocity for all three compositions is shown in figure~\ref{fig:r_1.5_theta_25_instant}g. 
We observe a rapid increase in $v_{avg}$ at early times followed by a slow increase afterwards. 
In addition, $v_{avg}$ at steady state decreases with increase in the composition of large species $f^T_L$. These effects are captured very well using the continuum model due to inter-coupling of rheology and segregation. The predicted instantaneous concentration profiles corresponding to $f^T_L = 0.25$ and $f^T_L = 0.75$ also show a very good match (see figure~\ref{fig:r_1.5_theta_25_instantfl_25_75} of the supplementary material) with DEM data.

\begin{figure}[h]
    \centering
    \begin{tikzpicture}
        \node[anchor=south west] (base4) at (4.2,0) 
            {\includegraphics[scale=0.32, trim=0 0 0 35, clip]{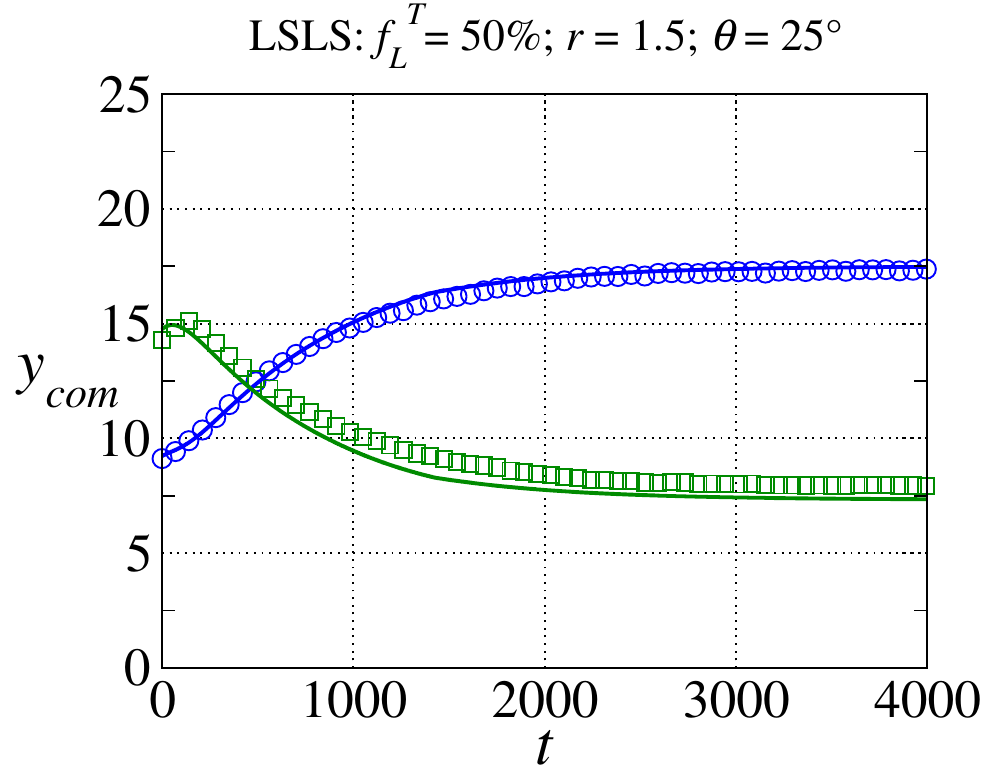}
            \put(-45,30){\textcolor{OliveGreen}{Small}}\put(-45,85){\textcolor{Blue}{Large}}
            }; 
        \node at (4.4,3.9) {(b)};     
        \node[anchor=south west] at (5.35,4) 
            {\includegraphics[scale=0.12, trim=200 0 30 45, clip]{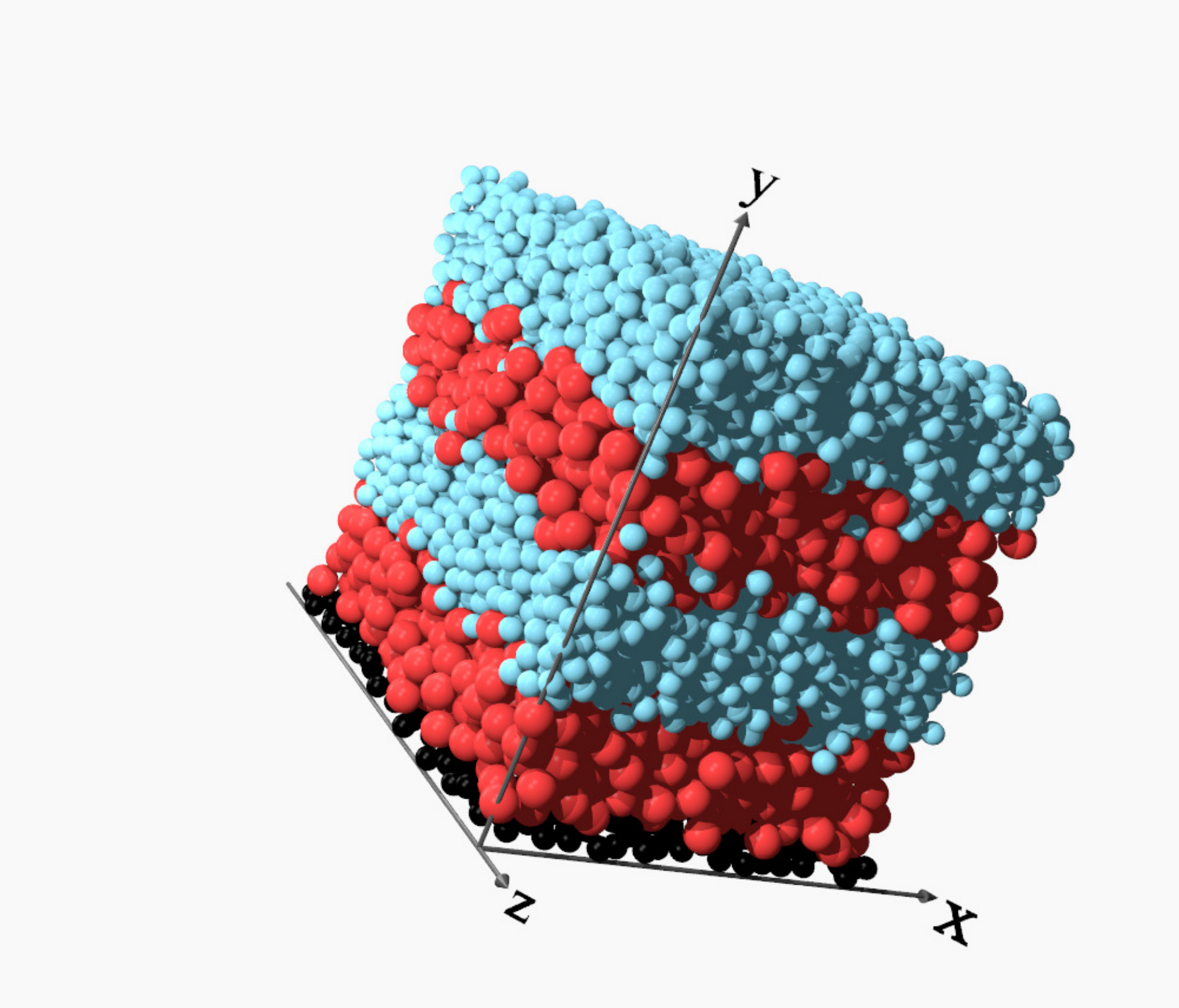}}; 
            \node at (5.6,7.2) {(a)};     
     \node[anchor=south west] at (10.5,3.5)
            {
             \includegraphics[scale=0.30]{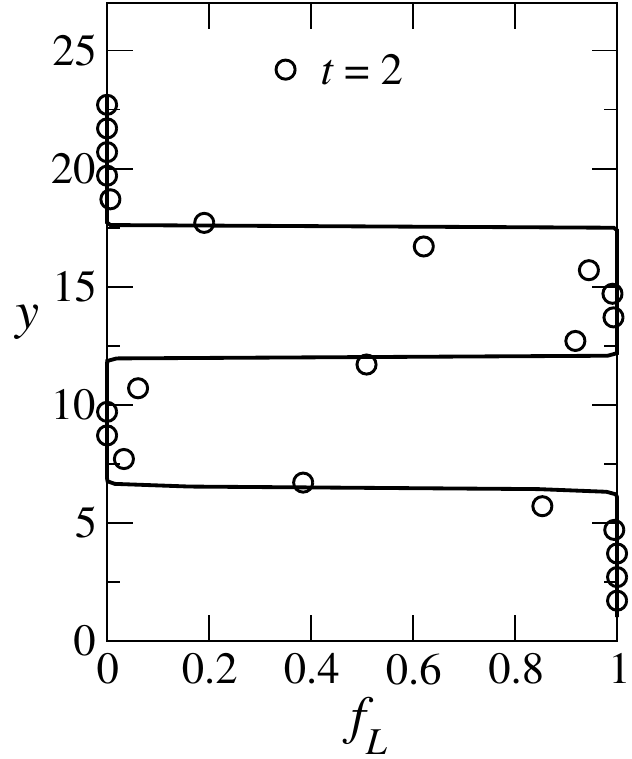}
            };
             \node at (10.35,7.5) {(c)}; 
         \node[anchor=south west] at (14.3,3.5)
            {
             \includegraphics[scale=0.30]{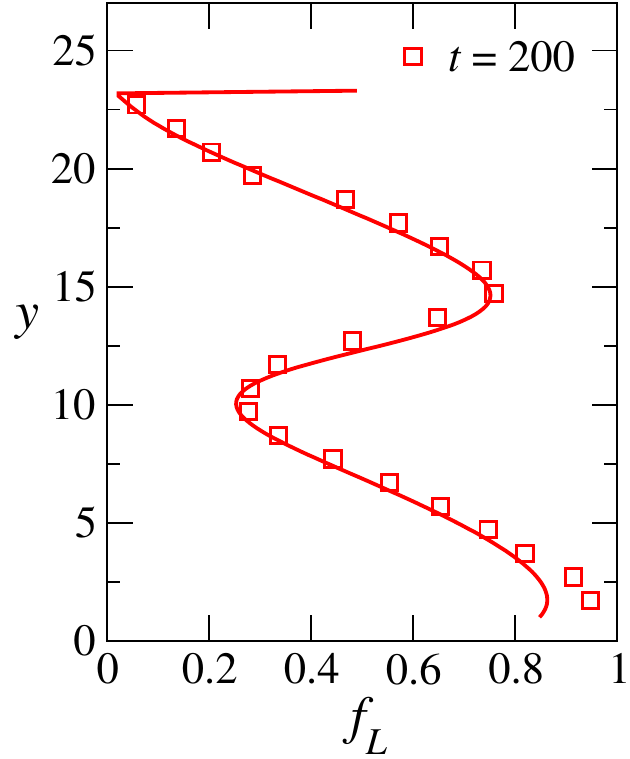}
            } ;   
             \node at (14.25,7.5) {(d)}; 
            \node[anchor=south west] at (10.5,-0.5)
            {
             \includegraphics[scale=0.30]{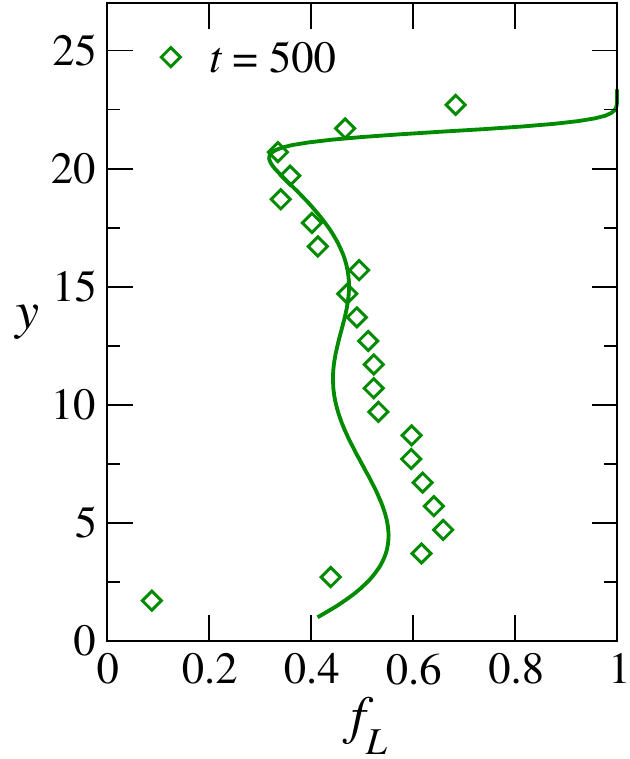}
            } ;  
            \node at (10.35,3.5) {(e)}; 
             \node[anchor=south west] at (14.3,-0.5)
            {
             \includegraphics[scale=0.30]{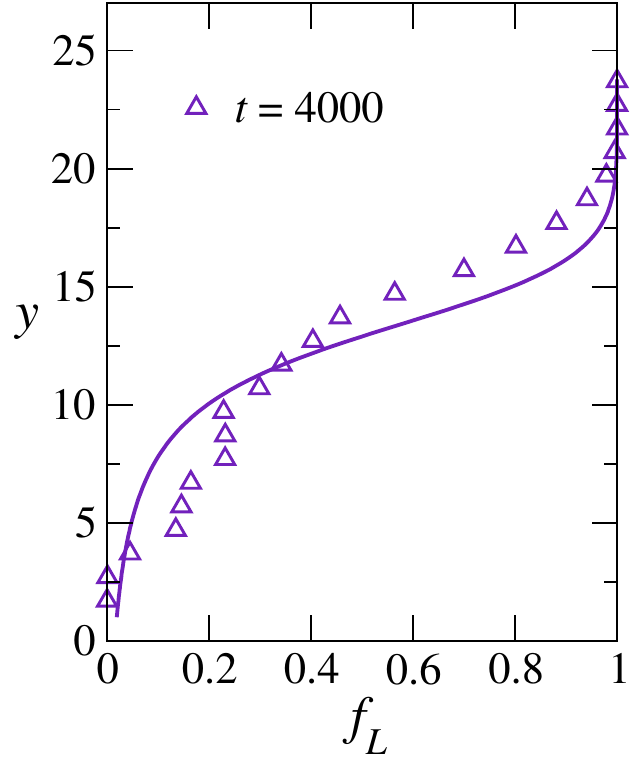}
            } ;   
            \node at (14.25,3.5) {(f)}; 
    \end{tikzpicture}
    \caption{(a) DEM snapshot showing a four layered initial configuration of an equal composition binary mixture. 
    (b) Time evolution of $y_{com}$ of both species. Instantaneous concentration profiles at different time for (c) $t = 2$, (d) $t = 200$, (e) $t = 500$, and (f) $t = 4000$ time units. Symbols represent the DEM data while solid lines correspond to model predictions.}
    \label{fig:Effect_conf_comp_LSLS}
\end{figure}
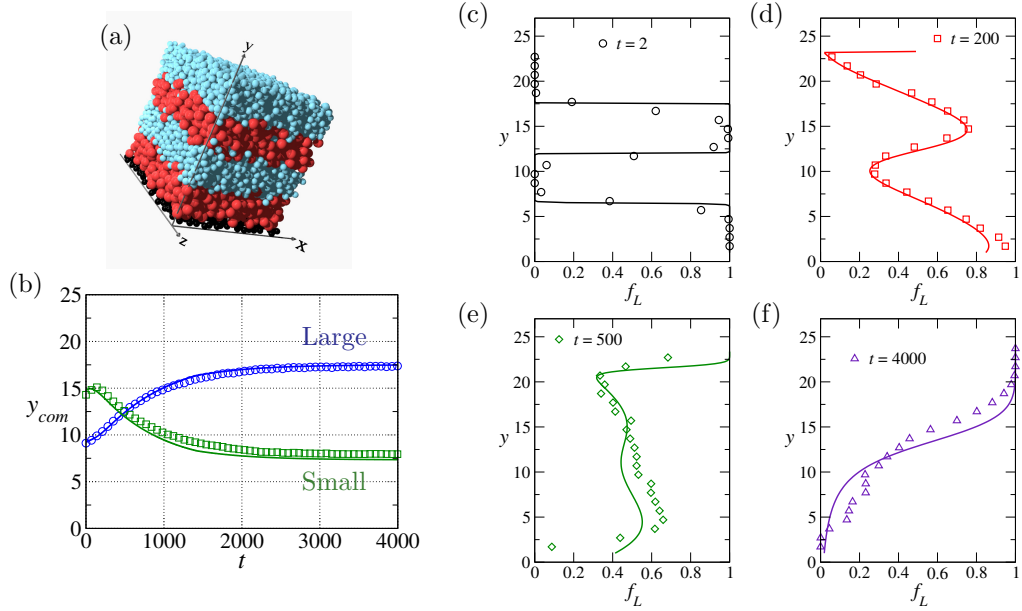
Next, we report the results for different initial configurations. Figure~\ref{fig:Effect_conf_comp_LSLS}a shows the DEM snapshot for a four layered initial configuration of an alternate arrangement of species for a $50\%–50\%$ binary mixture. The particle size ratio is $r = 1.5$ with a flowing layer height  $H \approx 25d$. 
Figure~\ref{fig:Effect_conf_comp_LSLS}b shows the evolution of the centre of mass for both large and small species. Evidently, the centre of mass of the species meet each other earlier compared to the two layer configuration shown in figure~\ref{fig:r_1.5_theta_25_instant}d. The continuum model is able to predict the $y_{com}$ evolution very well for this case as well. Figures~\ref{fig:Effect_conf_comp_LSLS}c – \ref{fig:Effect_conf_comp_LSLS}f show the instantaneous concentration profiles of large species at different times which is also very well captured by the continuum model. 
\begin{figure}[h]
    \centering
    \begin{tikzpicture}
        \node[anchor=south west] (base1) at (-10,0) 
            {\includegraphics[scale=0.26]{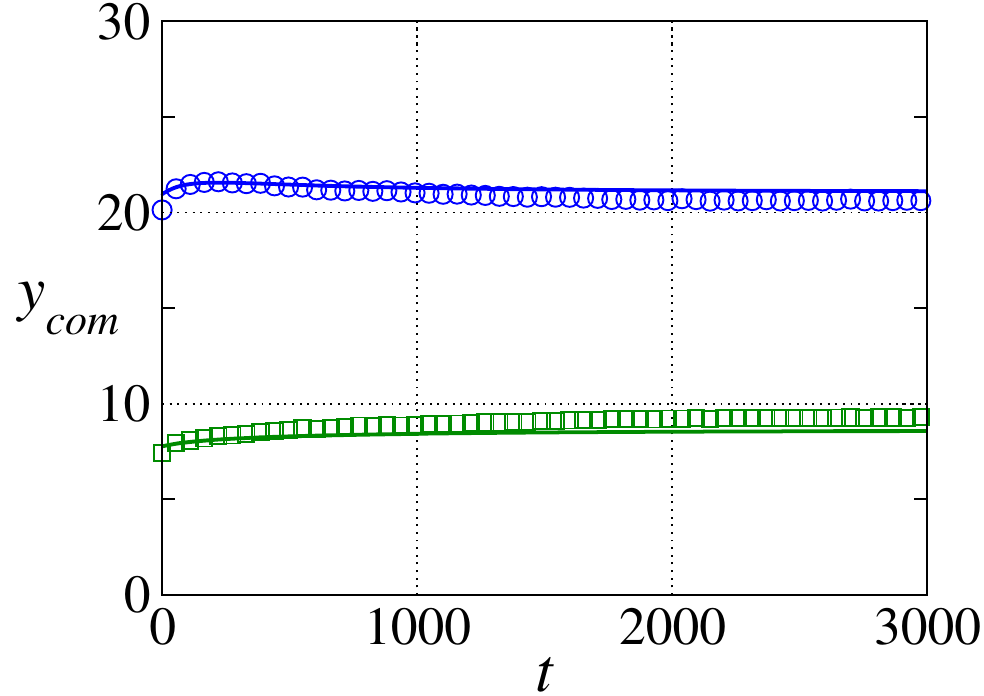}
            \put(-37,24){\textcolor{OliveGreen}{Small}}\put(-37,68){\textcolor{Blue}{Large}}
            };
        \node at (-10,3.2) {(a)};
        \node[anchor=south west] at (-9,2.4) 
            {\includegraphics[scale=0.0425, trim=200 0 30 0, clip]{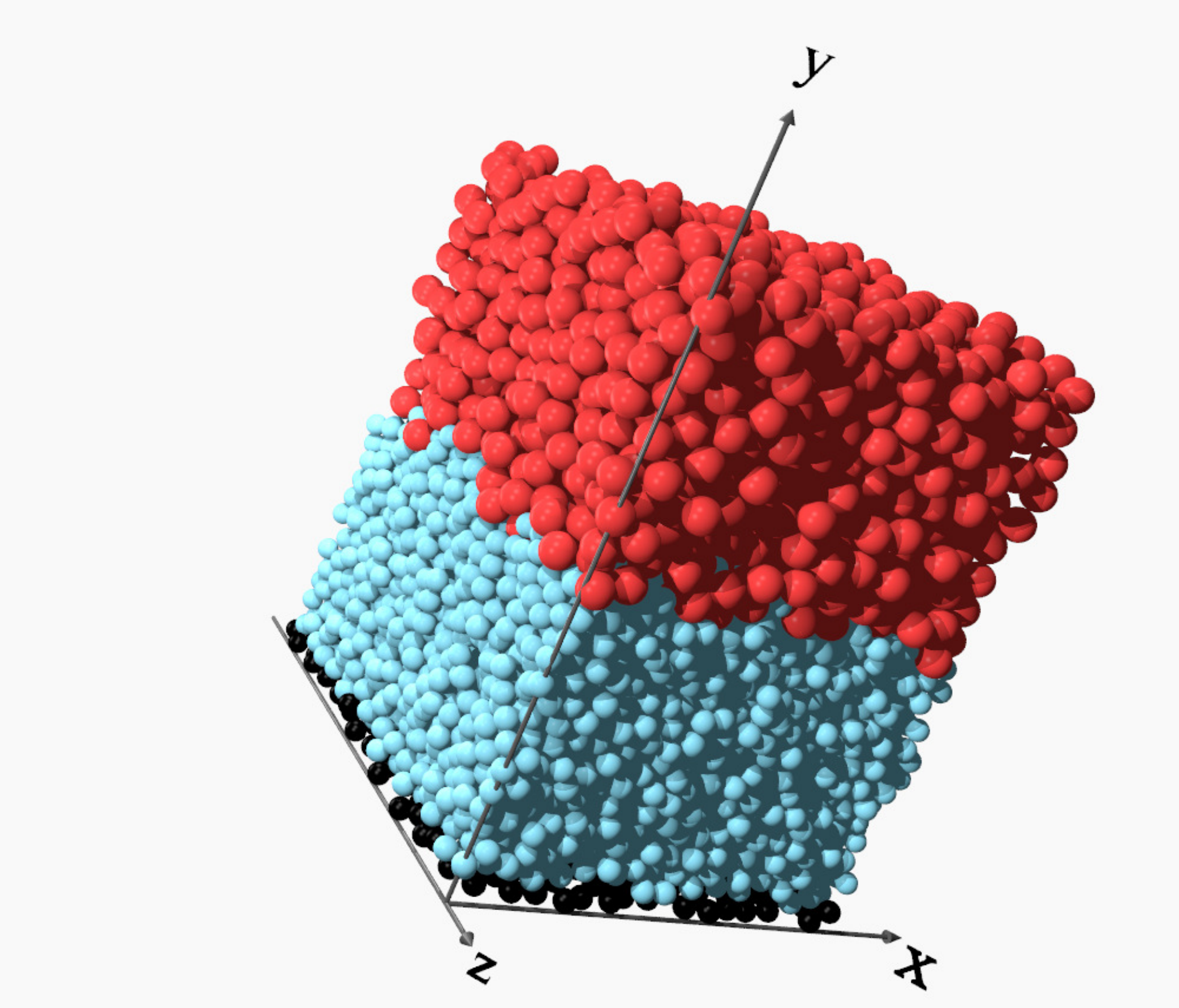}};
            \hspace{-0.6cm}
        \node[anchor=south west] (base2) at (-5,0) 
            {\includegraphics[scale=0.26]{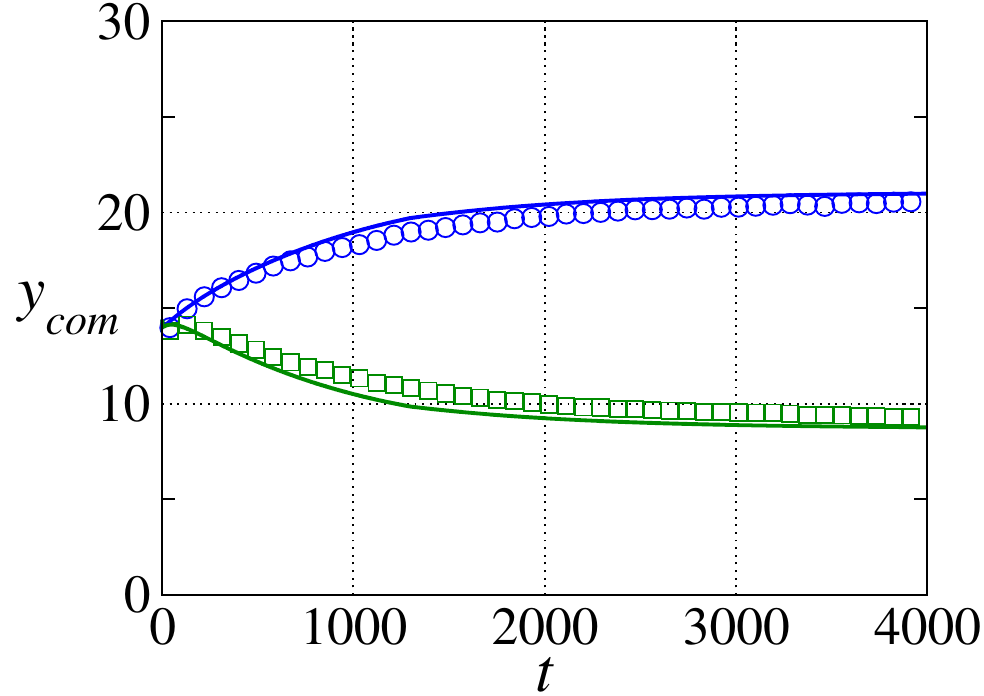}
             \put(-37,24){\textcolor{OliveGreen}{Small}}\put(-37,68){\textcolor{Blue}{Large}}
            };
        \node at (-4.8,3.2) {(b)};
        \node[anchor=south west] at (-4.2,2.4) 
            {\includegraphics[scale=0.0425, trim=200 0 30 0, clip]{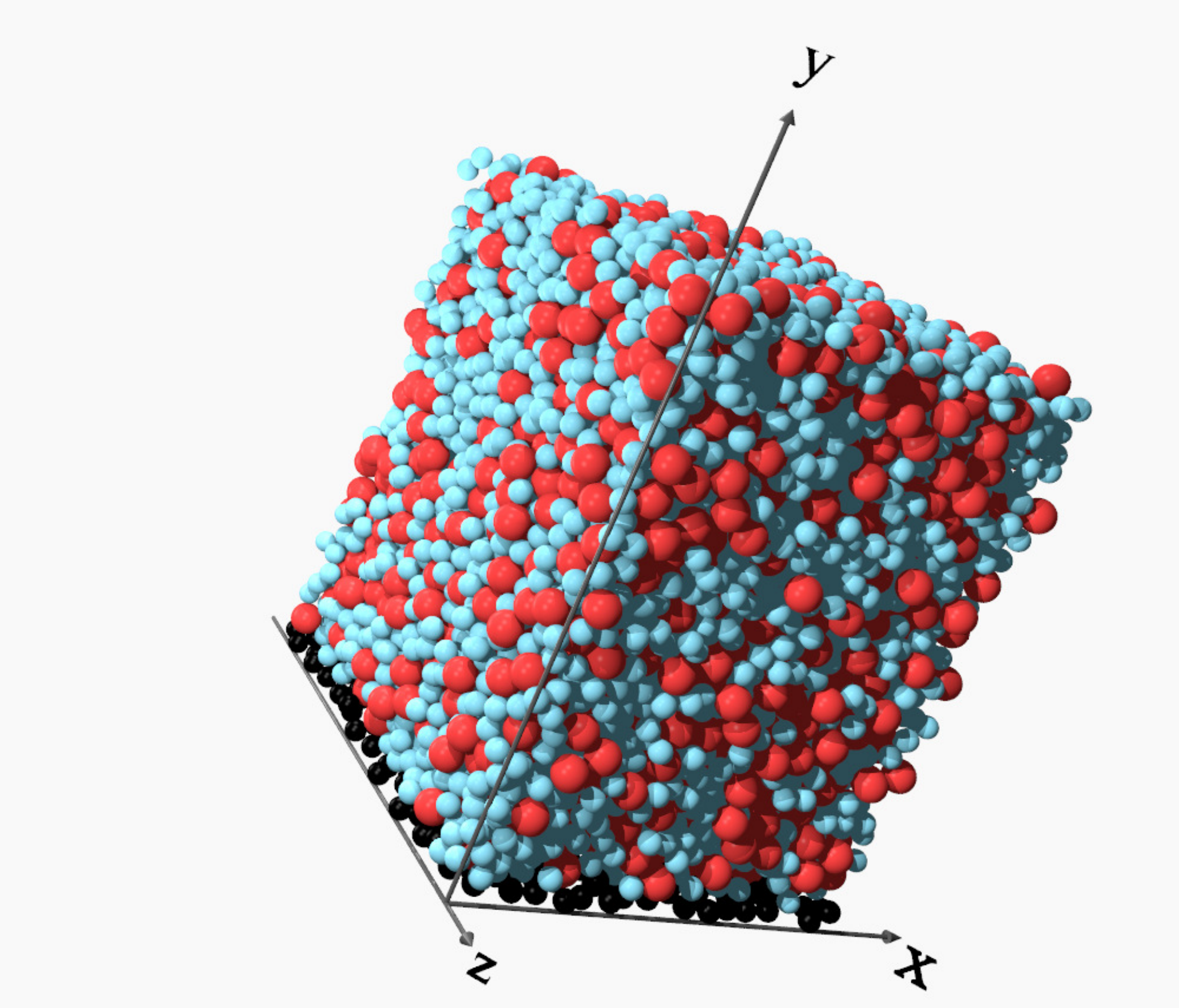}};
             \node[anchor=south west] (base3) at (-0.5,0) 
            {\includegraphics[scale=0.26, trim=0 0 0 30, clip]{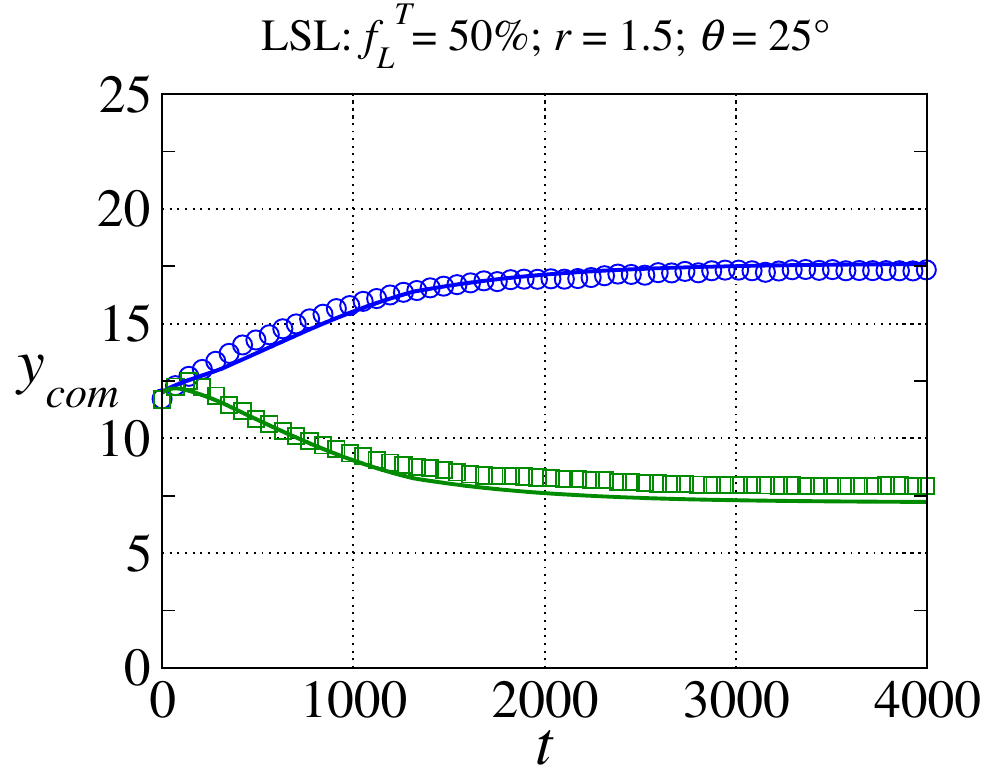}
             \put(-37,24){\textcolor{OliveGreen}{Small}}\put(-37,68){\textcolor{Blue}{Large}}
            };
        \node at (-0.3,3.2) {(c)};
        \node[anchor=south west] at (0.5,2.4) 
            {\includegraphics[scale=0.0425, trim=200 0 30 0, clip]{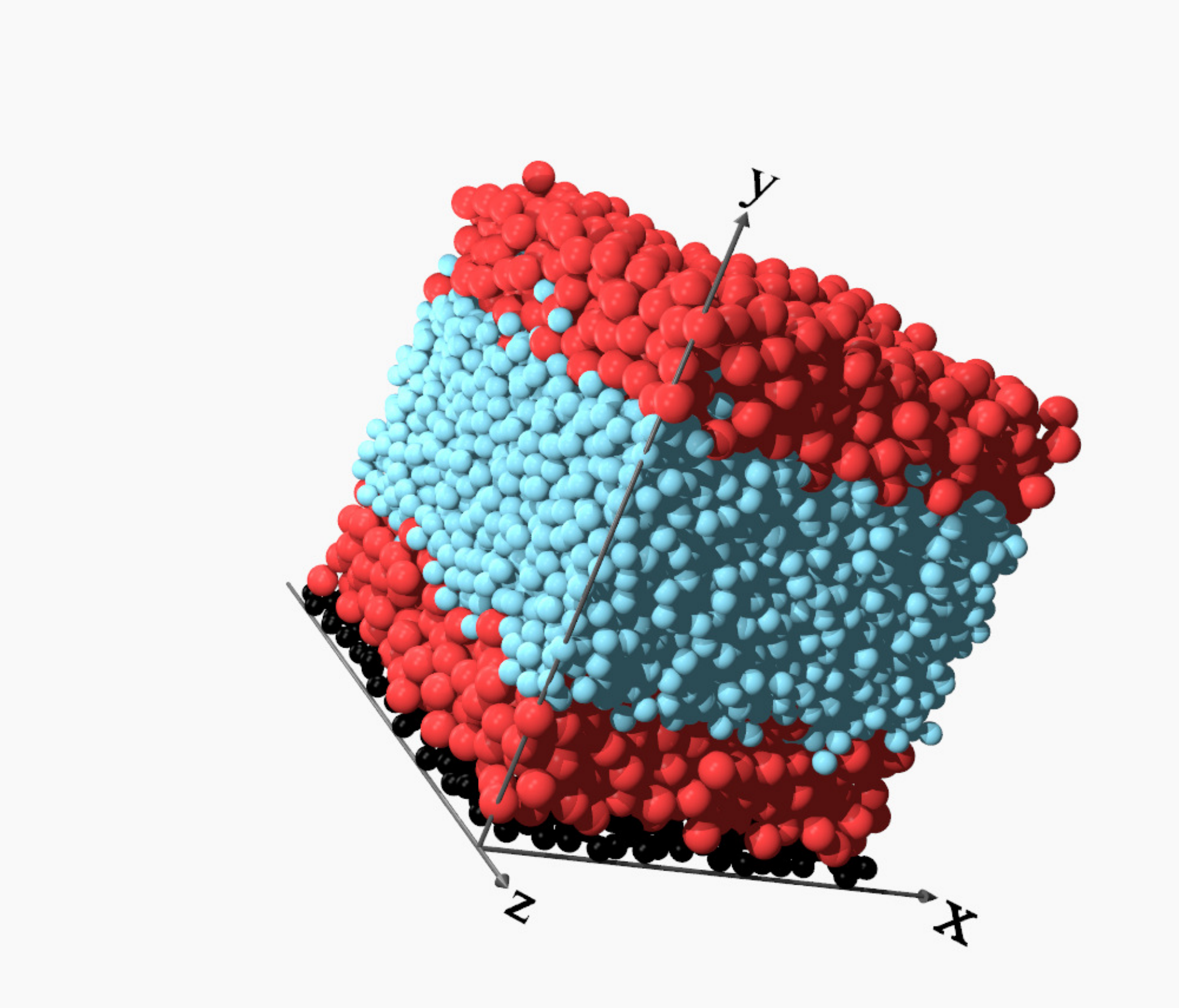}};
            \hspace{-0.6cm}
    \end{tikzpicture}
    \includegraphics[scale=0.32]{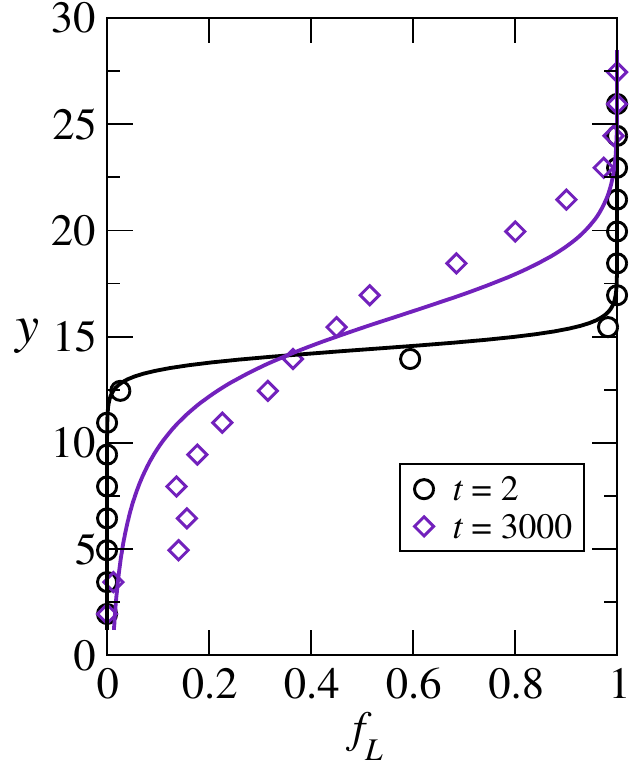}
    \put(-105,115){(d)}\quad \quad \quad 
    \includegraphics[scale=0.32]{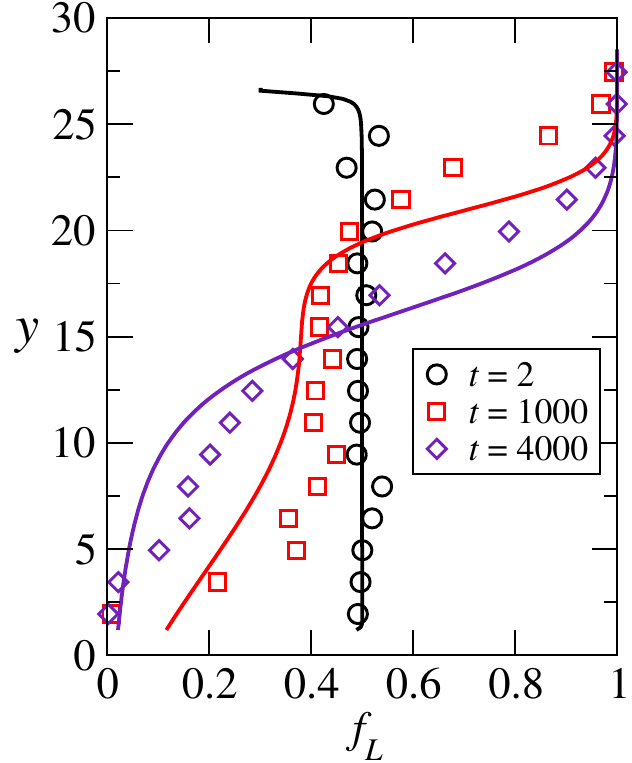}\put(-105,115){(e)} \quad \quad \quad
    \includegraphics[scale=0.32, trim=0 0 0 30, clip]{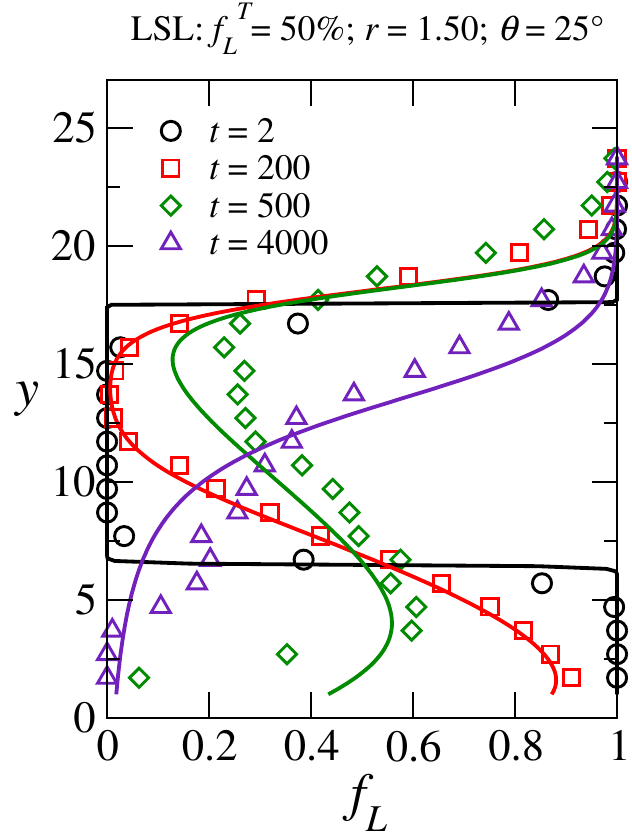}\put(-105,115){(f)} 
    \caption{Variation of $y_{com}$ with time for equal composition binary mixture having size ratio $r = 1.5$ flowing at an inclination angle of $\theta=  25 ^\circ$, starting from different initial configurations: (a) small near base, (b) uniformly mixed, and (c) small particles sandwiched between large particles. Corresponding instantaneous concentration profiles of large particles are shown in (d), (e), and (f) respectively. Symbols denote the DEM data and solid lines correspond to model predictions.}
    \label{fig:Effect_conf_comp}
\end{figure}
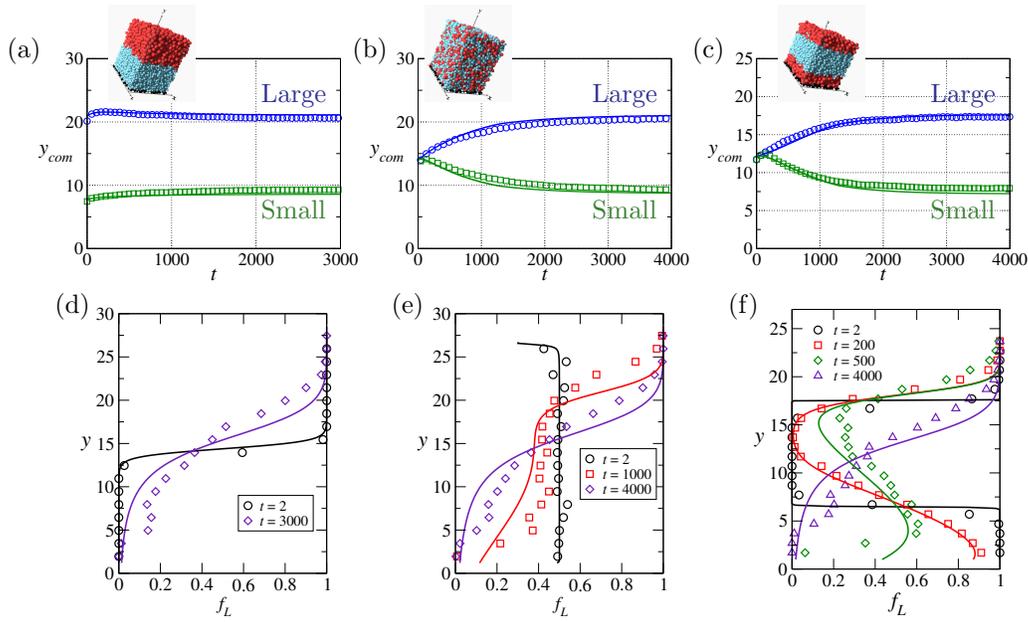
In order to test the efficiency of our model, we consider few other configurations for binary mixture with size ratio $r = 1.5$. Figure~\ref{fig:Effect_conf_comp} shows the evolution of the centre of mass of large and small species  for three different initial configurations of an equal composition mixture. 
Figure~\ref{fig:Effect_conf_comp}a shows the results for small near base and figure~\ref{fig:Effect_conf_comp}b shows the results for uniformly mixed configuration for layer thickness $H\approx 30d$.
The corresponding DEM snapshots are shown in inset. Since the flow starts from an initially segregated state in figure~\ref{fig:Effect_conf_comp}a, two species inter-diffuse into each other. While the instantaneous concentration profiles show this inter-diffusional mixing of the species clearly (see figure~\ref{fig:Effect_conf_comp}d), the centre of masses of the species seem to vary little and remain nearly at the same $y$ location. 
The $y_{com}$ of two species start from the identical $y-$ position in figure~\ref{fig:Effect_conf_comp}b because the particles are initially uniformly mixed.
As time progresses, large (blue) particles rise toward the surface, and smaller (green) particles settle near the base, giving rise to the increasing (decreasing) height of the large (small) species centre of mass.  
As expected, the steady state value of $y_{com}$ for both species in figures~\ref{fig:Effect_conf_comp}a and~\ref{fig:Effect_conf_comp}b are identical to $y_{com}$ value for large near base configuration, shown in the figure~\ref{fig:r_1.5_theta_25_instant}d.
Figure~\ref{fig:Effect_conf_comp}c shows the variation of $y_{com}$ corresponding to small particles sandwiched between large particles having size ratio $r = 1.5$ with flowing layer thickness $H  = 25d$. 
The predicted instantaneous concentration profiles for all three initial configurations show good agreement with the DEM data, as illustrated in figures~\ref{fig:Effect_conf_comp}d, \ref{fig:Effect_conf_comp}e, and \ref{fig:Effect_conf_comp}f, corresponding to the initial configurations shown in inset of figures~\ref{fig:Effect_conf_comp}a, \ref{fig:Effect_conf_comp}b, and \ref{fig:Effect_conf_comp}c, respectively.

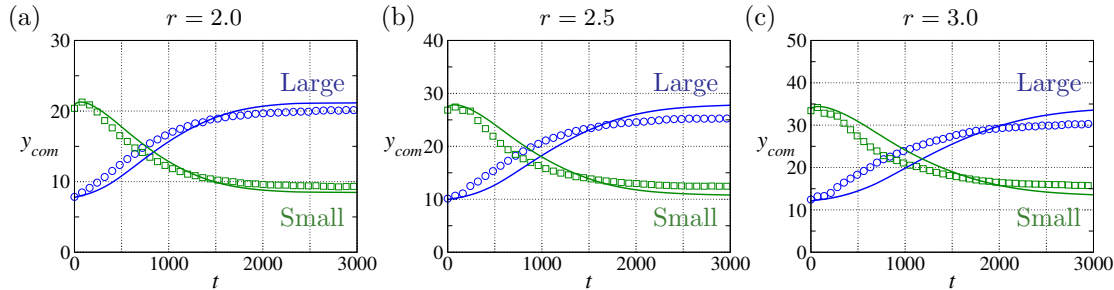
\begin{figure}[h]
    \centering
     \includegraphics[scale=0.29, trim=0 0 0 35, clip]{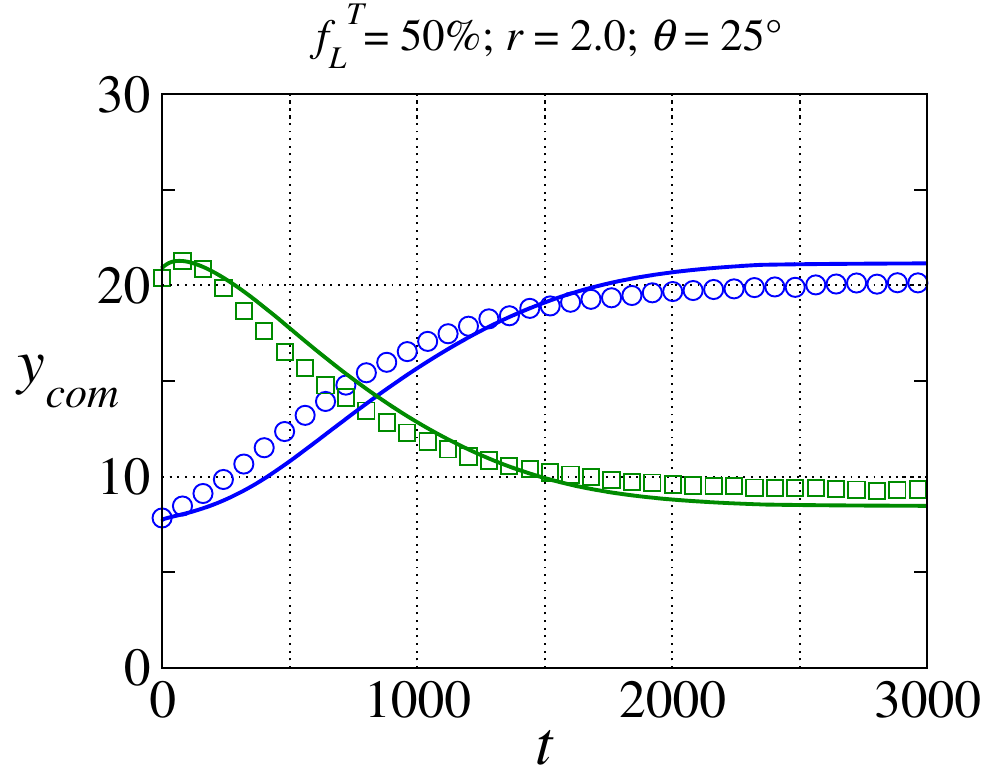}\put(-140,100){(a)}\put(-80,100){\small $r = 2.0$} \put(-37,24){\textcolor{OliveGreen}{Small}}\put(-37,75){\textcolor{Blue}{Large}}
    \includegraphics[scale=0.29, trim=0 0 0 35, clip]{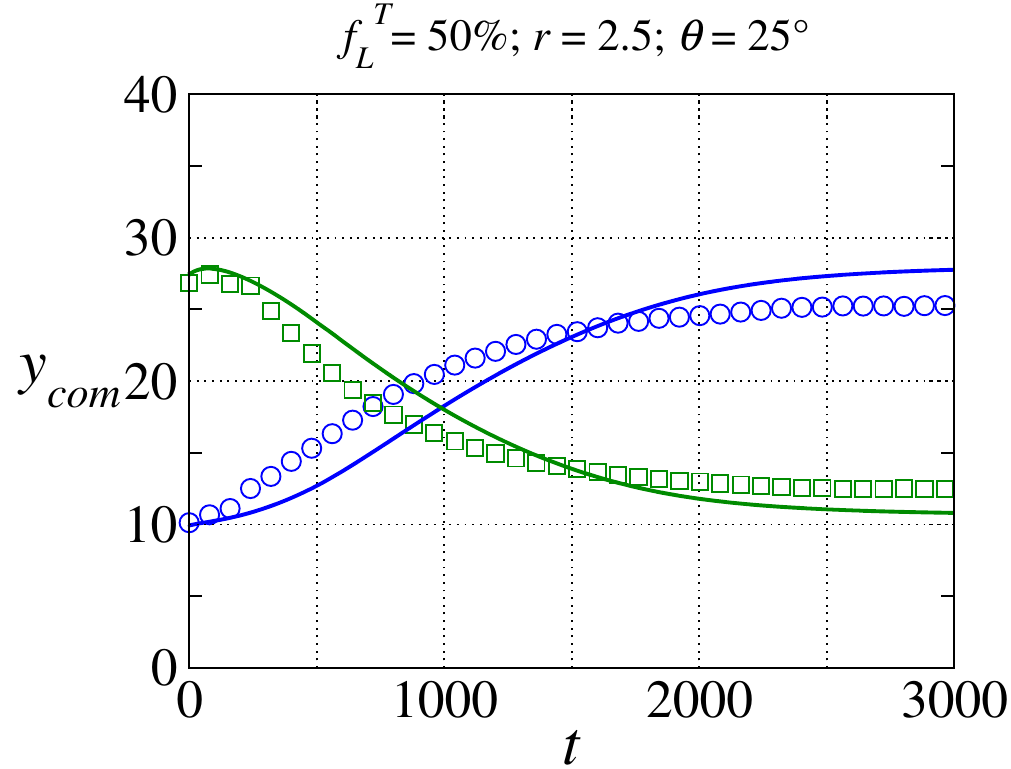}\put(-140,100){(b)}\put(-80,100){\small $r = 2.5$} \put(-37,24){\textcolor{OliveGreen}{Small}}\put(-37,75){\textcolor{Blue}{Large}}
    \includegraphics[scale=0.29, trim=0 0 0 35, clip]{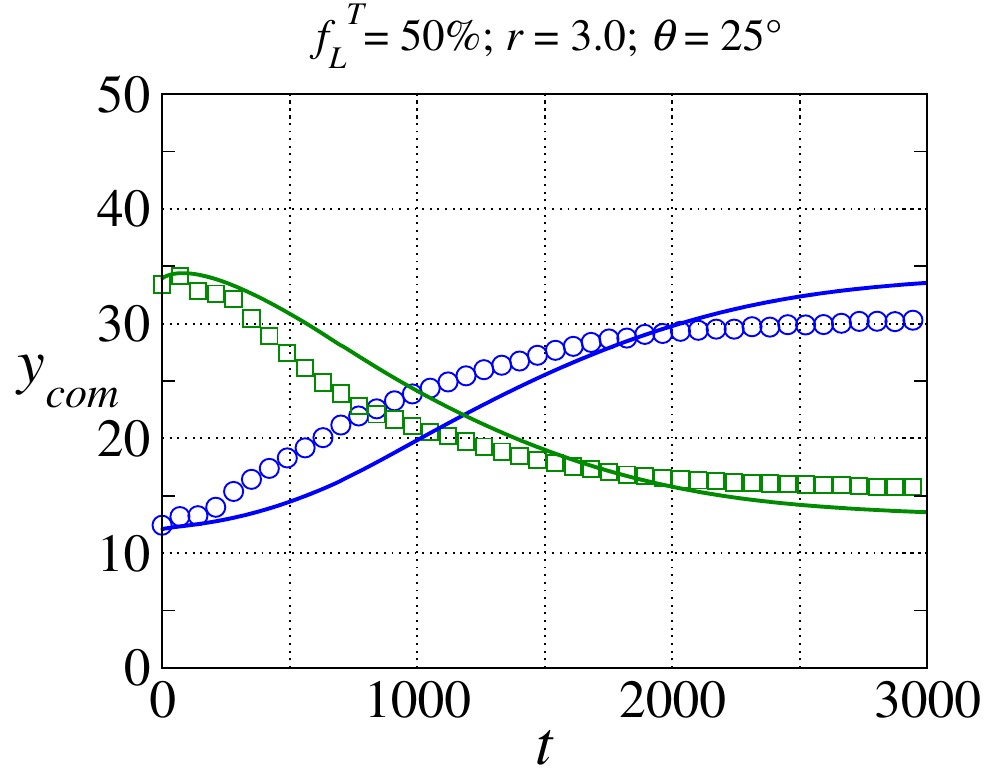}\put(-140,100){(c)}\put(-80,100){\small $r = 3.0$} \put(-37,24){\textcolor{OliveGreen}{Small}}\put(-37,75){\textcolor{Blue}{Large}}
    \caption{Evolution of $y_{com}$ for binary mixture of size ratios (a) $r = 2.0$, (b) $r = 2.5$, and (c) $r = 3.0$. The composition of the mixture in all the cases is $f^T_L = 0.5$ and the inclination angle is $\theta = 25^\circ$.}
    \label{fig:theta_size_effect}
\end{figure}

We now report the effect of size ratio of the species on the evolution of center of mass of the two species. 
Figures~\ref{fig:theta_size_effect}a, \ref{fig:theta_size_effect}b, and~\ref{fig:theta_size_effect}c show the $y_{com}$ evolution for equal composition binary mixture having size ratios $r = 2.0$, $r = 2.5$, and $3.0$, respectively. 
Note that the flowing layer height for different sizes are different. For larger size ratios of $2.5$ and $3.0$, slight deviation of the model predictions from DEM data are observable. Figure~\ref{fig:size_effect_r_1.25_1.5_2_2.5_3} of the supplementary material shows the data for all the different size ratios considered in this study in a slightly different manner. Evidently, the continuum model predictions of $y_{com}/h_0$ for size ratios upto $2$ are in excellent agreement with DEM simulations. 
For size ratios $r>2$, an interesting observation from the DEM simulations is that the evolution of small particles appears nearly identical across different cases, whereas the behavior of large particles varies noticeably (figure~\ref{fig:size_effect_r_1.25_1.5_2_2.5_3}b). Our model captures the differences in the large particle dynamics well, but it does not reproduce the nearly identical evolution of the scaled center of mass position for the small particles (figure~\ref{fig:size_effect_r_1.25_1.5_2_2.5_3}c). However, the model does reasonably capture the $y_{com}$ profiles for mixtures with compositions of $f^T_L = 0.25$ and $f^T_L = 0.75$ at $r = 2.5$ and $3.0$ (see figure~\ref{fig:flt_0.25_0.75_r_2.5_3} of the supplementary material).

To summarize, we developed a one-dimensional continuum model to predict evolution of flow properties for bi-dispersed granular mixtures flowing over a periodic chute. The model incorporates the particle force-based segregation model~\cite{yennemadi2023drag} in the time-dependent segregation-diffusion equation along with an inertial number based mixture rheological model~\cite{tripathi2011rheology} in the momentum balance equations. 
Contrary to many other approaches~\cite{Fan2015,deng2019modeling} that require apriori knowledge of flow kinematics for solving the segregation-diffusion equation, our model solves the momentum balance equations to determine the evolving flow kinematics and accounts for the inter-coupling of the rheology with segregation. The model is able of predict various flow properties such as the velocity, shear stress, inertial number, and concentration profiles for various initial configurations of the two species. The model predictions are found to be in good agreement with the DEM simulations. The model also captures the effect of mixture composition, and the large to small particle size ratios for different flowing layer thicknesses. We wish to emphasize that by using a periodic chute, we are essentially simulating a fixed mass system. Such fixed mass systems have been used in other works as well \citep{marks_rognon_einav_2012,trewhela2024segregation,singhandhennan2024continuum} and are conceptually similar to experimental systems like oscillating shear flow or Couette flow. Hence, our periodic chute results are qualitatively different from those expected in the case of flow over a long chute, where the mass flux of the two species is constant at different locations along the chute. More specifically, periodic chute results at early (later) times must not be considered similar to those observed for upstream (downstream) chute positions. The continuum model presented here can be easily extended to account for the presence of the convective term and would require solving the variation of properties along $x-$ direction as well. Such an extension of our continuum model to predict the evolution of the mixture properties while flowing over a long chute is currently under investigation. In future, it will also be worth investigating whether the approach can be generalized to polydisperse mixtures. This study paves the way for predicting combined size and density segregation using particle force-based segregation models.


\section*{ACKNOWLEDGEMENTS}
AT gratefully acknowledges the financial support provided by the Indian Institute of Technology Kanpur via the initiation grant IITK/CHE/20130338. A.T. and S.K. gratefully acknowledge the funding support provided to S.K. by the Prime Minister’s Research Fellowship (Government of India) grant.

\section*{DATA AVAILABILITY}
The data that support the findings of this study are available from the corresponding author upon request.

\bibliographystyle{apsrev4-2}
\bibliography{aipsamp}%

\newpage
\section*{Supplementary Material}
\beginsupplement
\renewcommand{\thesection}{SM\arabic{section}} 
\setcounter{section}{0} 
\setcounter{linenumber}{0} 

\section{$\mu - I$ and $\phi - I$ rheology}
\label{sec:mu_I_phi_I}
The shear stress ($\tau_{yx}$) and pressure ($P$) in dense granular flows are related by means of the effective friction coefficient dependent on the inertial number $\mu(I_{mix}) = |\tau_{yx}|/ P$. 
The JFP model \cite{jop2006constitutivenature} empirically relates the variation of $\mu$ with $I$ as 
\begin{equation}
\mu(I_{mix}) = \mu_{s}+\frac{\mu_{m}-\mu_{s}}{1+I_{0}/I_{mix}}, 
\label{eq_5:mu_I_JFP}
\end{equation}
where, $\mu_s$, $\mu_m$, and $I_0$ are the rheological model parameters. The expression of the inertial number for the granular mixture having identical density ($\rho_{mix} = \rho_p$) and different size ($d_{mix} = \sum_i d_i f_i$) particles is given by $ I_{mix}=|\dot\gamma| d_{mix} /\sqrt {P/\rho_p}$, where $\dot \gamma =dv_x/dy$ is the local shear rate. As a first approximation, the local solids fraction $\phi$ can be obtained using the dilatancy law for monodisperse grains using the generalized inertial number \cite{tripathi2011rheology} as 
\begin{equation}
    \phi_{th}(y,t)=\phi_{max}- \beta I_{mix}. 
    \label{eq:packing_fraction}
\end{equation}
The parameters $\phi_{max}$ and $\beta$ of the dilatancy law, obtained from the DEM simulation data, are the same as those used for the monodisperse case.
The values of these parameters are taken from \cite{tripathi2013density,sahu_kumawat_agrawal_tripathi_2023} and are reported in table~\ref{tab:rheopar}.
\begin{table}[h]
\centering
  \begin{center}
    \begin{tabular}{cccccc}
    \hline
    $\mu_{s}$ & $\mu_{m}$ & $I_{o}$  &$\phi_{max}$ & $\beta$ \\
    \hline
    $\tan(20.16^{\circ})$ & $\tan(37.65^{\circ})$ & 0.434 & 0.59 & 0.16 \\
    \hline
\end{tabular}
    \caption{Values of the rheological model parameters (same as in \cite{tripathi2011rheology}) used in this study.}
    \label{tab:rheopar}
  \end{center}
\end{table}

Following \cite{tripathi2011rheology}, we observe that the solids fraction for binary-size mixtures is relatively higher than that for monodisperse mixtures and increases with the size ratio. We, therefore, adopt a modified form of the dilatancy law following their approach and use
\begin{equation}
     \phi(y,t)  = g(f_L) \phi_{th}(y,t)
\end{equation}
where, $\phi_{th}(y,t)$ is calculated using equation~\ref{eq:packing_fraction}. $g(f_L)$ is obtained by fitting a cubic function to the $\phi - f_L$ data obtained from DEM simulations for different size ratio mixtures. Thus $g(f_L)$ is expressed as a cubic function of the large species concentration ($f_L$)
\begin{equation}
    g(f_L) = 1 - (a_0 + a_1)f_L +a_0f_L^2 + a_1f_L^3,
\end{equation}
where $a_0$ and $a_1$ are fitting that depend on the size ratio. This functional form ensures that $g(f_L) = 1$ in the limiting cases of $f_L = 0$ and $f_L = 1$, so that the solids fraction of the mixture remains consistent with the solid fraction of monodisperse grains in the case of only one type of particles. The values of $a_0$ and $a_1$ for different size ratios are listed in table~\ref{tab:g_f_parameters}.


\begin{table}[h]
    \centering
    \begin{tabular}{cccccc}
       \hline
        \textbf{Size ratio} ($r$) & 1.5 & 1.75 & 2.0 & 2.5 & 3.0 \\
       \hline
        $a_0$ & -0.06 & 0.01 & 0.08 & 0.20 & 0.40 \\
        $a_1$ & -0.05 & -0.13 & -0.25 & -0.41 & -0.59 \\
        \hline
    \end{tabular}
    \caption{Fitting parameters for solids fraction ($\phi$) versus concentration ($f_L$) for binary mixtures with size ratio $r$.}
    \label{tab:g_f_parameters}
\end{table}

\section{Numerical Solution Methodology}
\label{sec:NumericalMethod}
In order to predict the evolution of the properties of the mixture with time, the initial and boundary conditions are required for solving segregation-diffusion equation (Eq~\ref{eq:Conc_pde}) and momentum balance equation (Eq.~\ref{eq:mombal_x}).
The initial and boundary conditions for solving equation (Eq~\ref{eq:Conc_pde}) are given as 
\begin{equation}
\begin{split}
     f_{L}(y,0) = f_{L,ini}(y); \\
     J_{L}^{S}(0,t) + J_{L}^{D}(0,t) = 0; \\
     J_{L}^{S}(h,t) + J_{L}^{D}(h,t) = 0; 
    \end{split}
\label{eq:IC_BC_concentration_multi}
\end{equation}
where, $f_{L,ini}(y)$ is the initial concentration profile of the large species. In case of starting from a uniformly mixed configuration, $f_{L,ini}$ is equal to the total composition of that species in the mixture, that is, $f^T_L$. For the cases where the mixture is nearly segregated, $f_{L,ini}(y)$ is given by a step function. In order to enable comparison with our DEM simulation results, the initial concentration profile is approximated by fitting a sigmoid function to the concentration profile obtained from the DEM simulation data. The boundary conditions correspond to no net mass flow in the $y$ direction at the free surface ($y=h$) and the base ($y=0$) and state that the sum of segregation and diffusion fluxes will be zero at extreme values of $y$.

In our simulations, the flow is started from an initial condition of zero velocity and we incorporate a rough, bumpy base to ensure the no-slip boundary condition at the base. The initial and boundary conditions for solving unsteady momentum balance equation (eq~\ref{eq:mombal_x}) for this situation are given as follows: 
\begin{equation}
   \begin{split}
        IC: v_{x}(y,0) = 0,\\
        BC1: v_{x}(0,t) = 0,\\
        BC2: \tau_{yx}(h,t) =0,
   \end{split} 
    \label{eq:IC_BC_momentum}
\end{equation}
where BC2 corresponds to the zero stress condition at the free surface.
\textcolor{black}{The species concentration and velocity fields are obtained by solving the segregation-diffusion equation (Eq~\ref{eq:Conc_pde}) and momentum balance equation (Eq~\ref{eq:mombal_x}) simultaneously using the PDEPE solver along with the initial and boundary conditions given in equation~\ref{eq:IC_BC_concentration_multi} and equation~\ref{eq:IC_BC_momentum}, respectively. 
The PDEPE solver uses numerical discretization to approximate the derivatives by dividing the domain into finite spatial and temporal grids. A total of $N = 200$ spatial grids along the $y$ direction are used which corresponds to a grid size of approximately one-eighth of the particle diameter. We choose the time step $\Delta t = 0.1$ dimensionless time units. This time step corresponds to approximately $10^{-3}s$ for 1 mm particle size for earth's gravitational acceleration. In order to facilitate comparison with DEM simulations which report properties averaged over two time units, the instantaneous properties from the continuum model predictions are also obtained by averaging $20$ snapshots over a time period of two time units.}

\section{DEM Simulation Methodology}
\label{sec:simulationMethod}
We numerically simulate spherical particles of different sizes and the same density ($\rho_p$) flowing down an inclined plane using the Discrete Element Method (DEM). The particles are assumed to be frictional (inter-particle friction coefficient $\mu_{pp} = 0.5$) and slightly inelastic (normal restitution coefficient $e = 0.88$). The simulation domain spans an area of $20d \times 20d$ in the streamwise ($x-$) and vorticity ($z-$) directions, with periodic boundary conditions imposed along both directions. The height of the simulation box in the $y$ direction varies with the size ratio, and is approximately $15 d_L$. At the start of the DEM simulation, particles are arranged in a cubic lattice to ensure no initial contact between adjacent particles. The particles are then allowed to settle under gravity at an inclination angle of $\theta = 0^\circ$. After settling, the inclination angle is increased to the desired value to initiate the flow. The linear spring and dashpot model is used to compute contact forces in both normal and tangential directions. The normal spring stiffness is $k_{n} = 2 \times 10^{5} mg/d$, while the tangential spring stiffness is $k_{t} = 2k_{n}/7$. More details on the contact force model and properties calculations are provided in \cite{tripathi2011rheology,tripathi2021size}. 
\begin{figure}[h]
    \centering
   \includegraphics[scale=0.12, trim=200 0 30 0, clip]{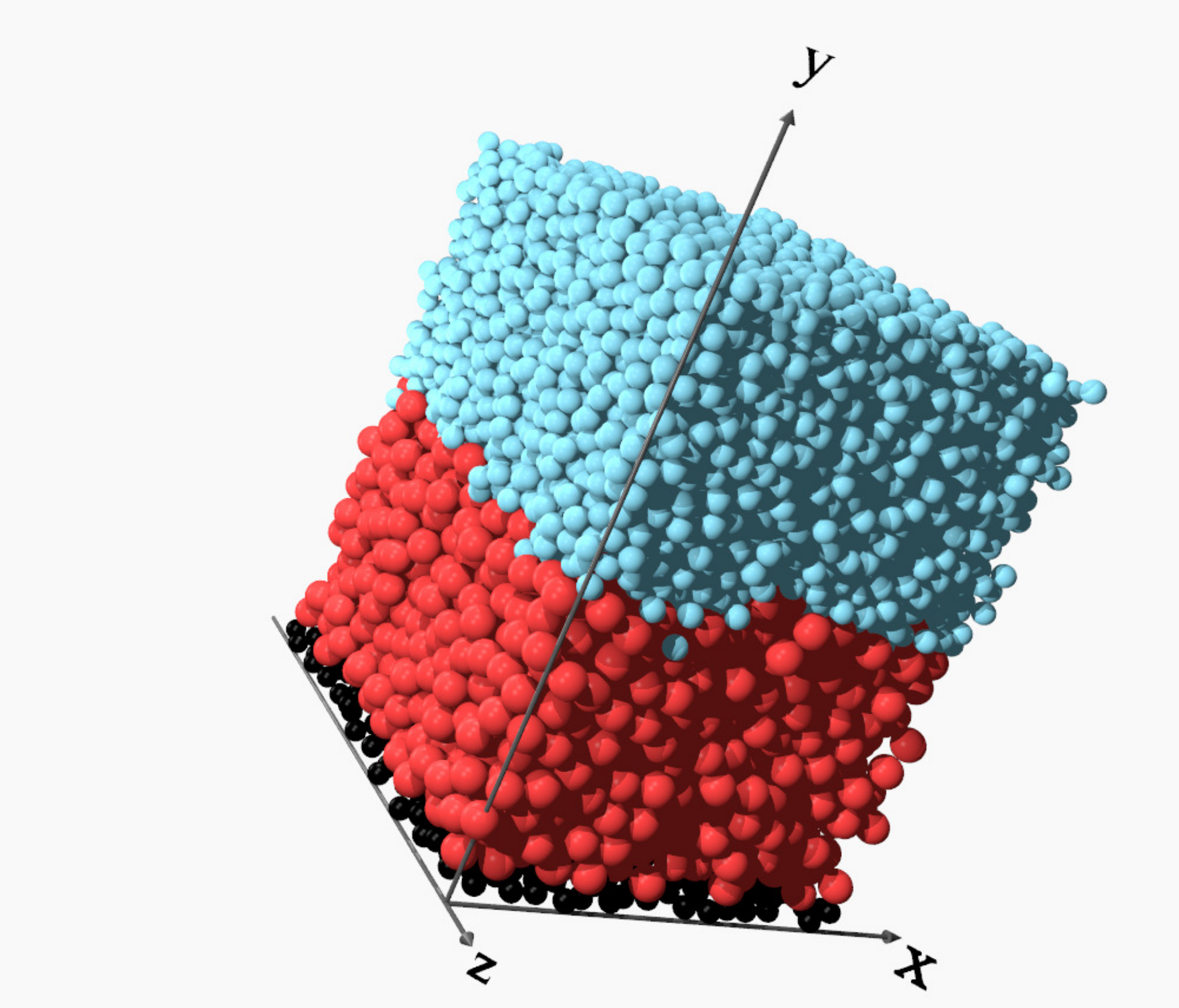}\put(-100,115){(a)}
    \includegraphics[scale=0.13, trim=200 0 30 0, clip]{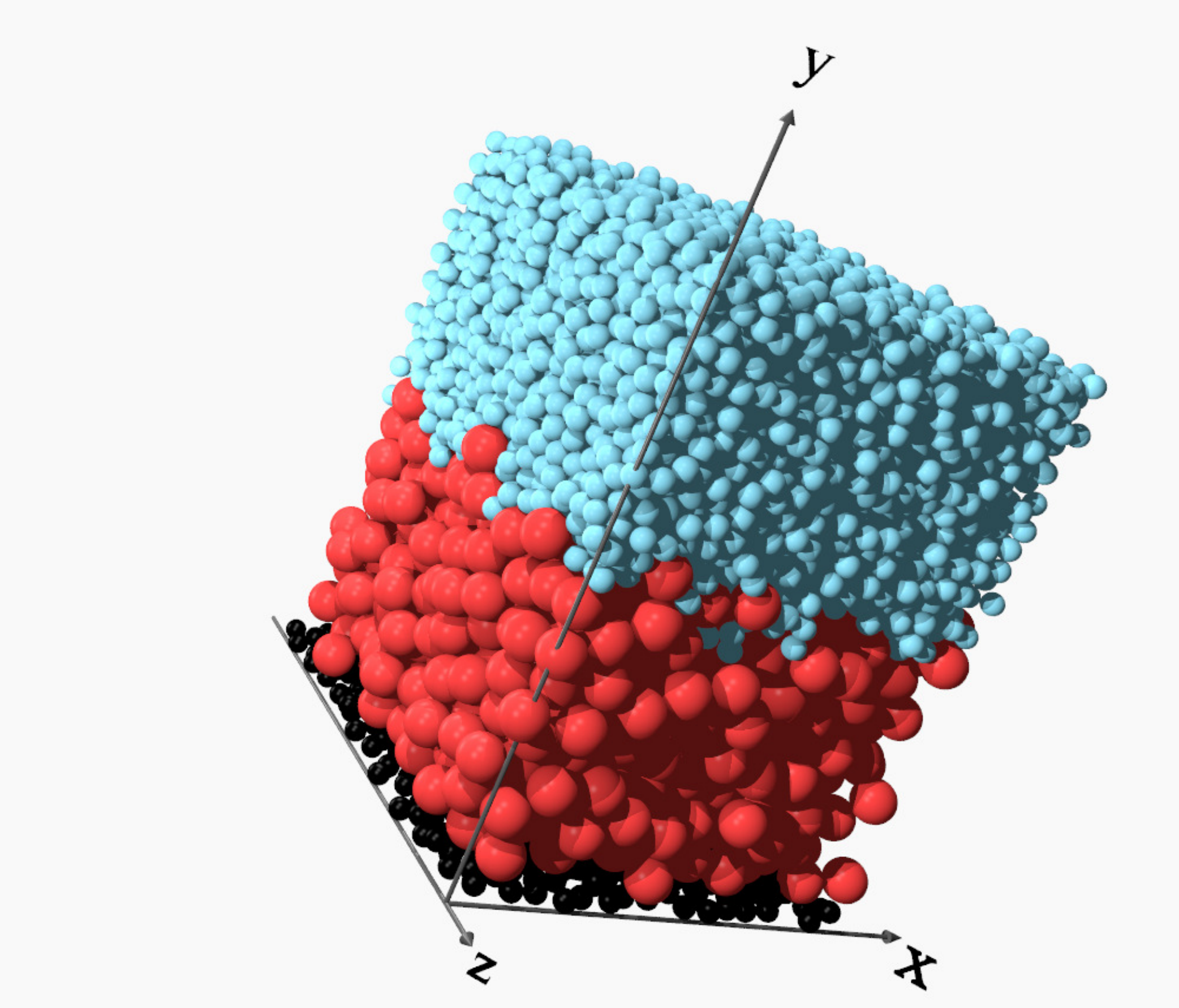}\put(-100,115){(b)}
     \includegraphics[scale=0.21, trim=200 110 190 0, clip]{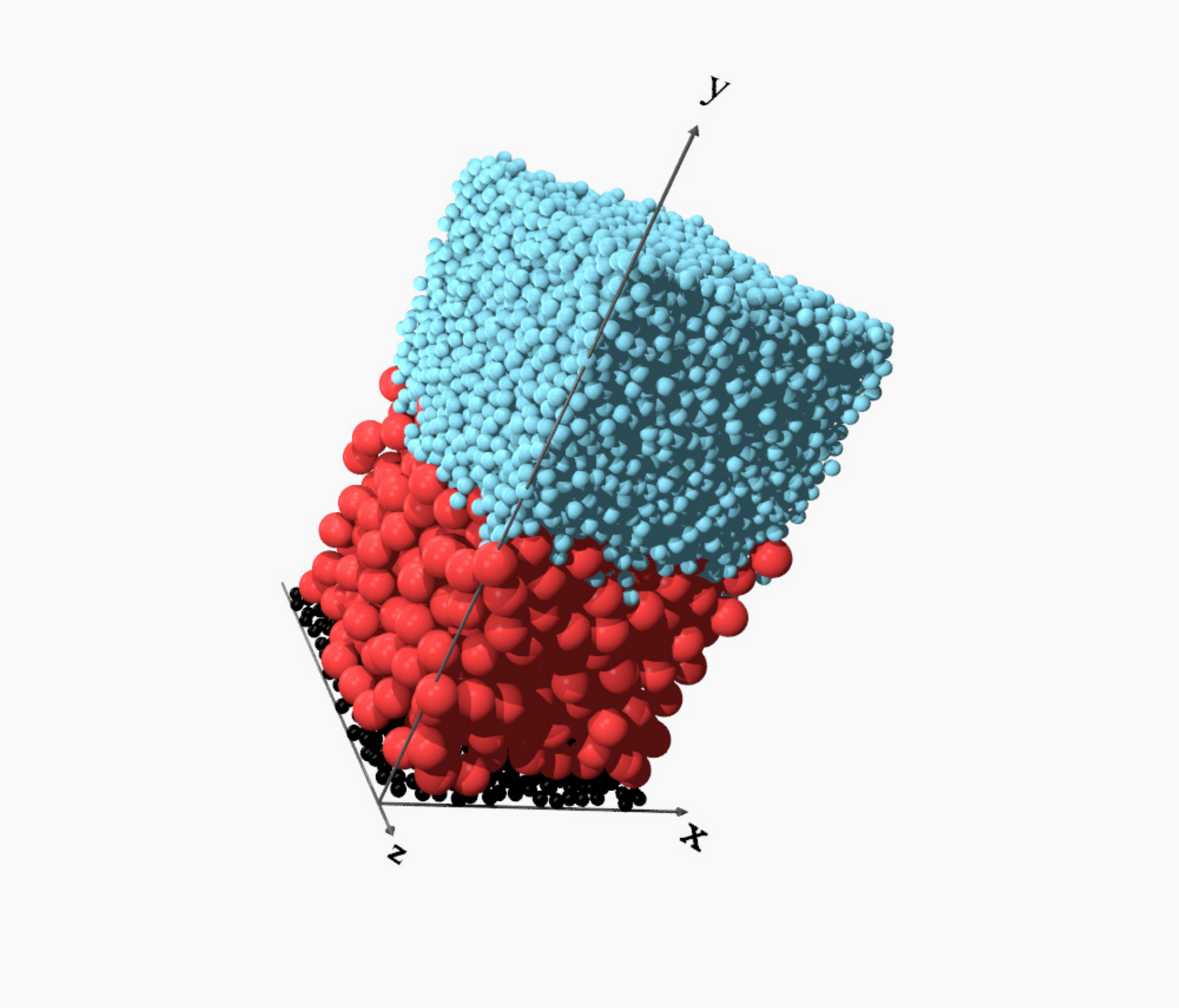}\put(-100,140){(c)}
      \includegraphics[scale=0.23, trim=200 80 190 0, clip]{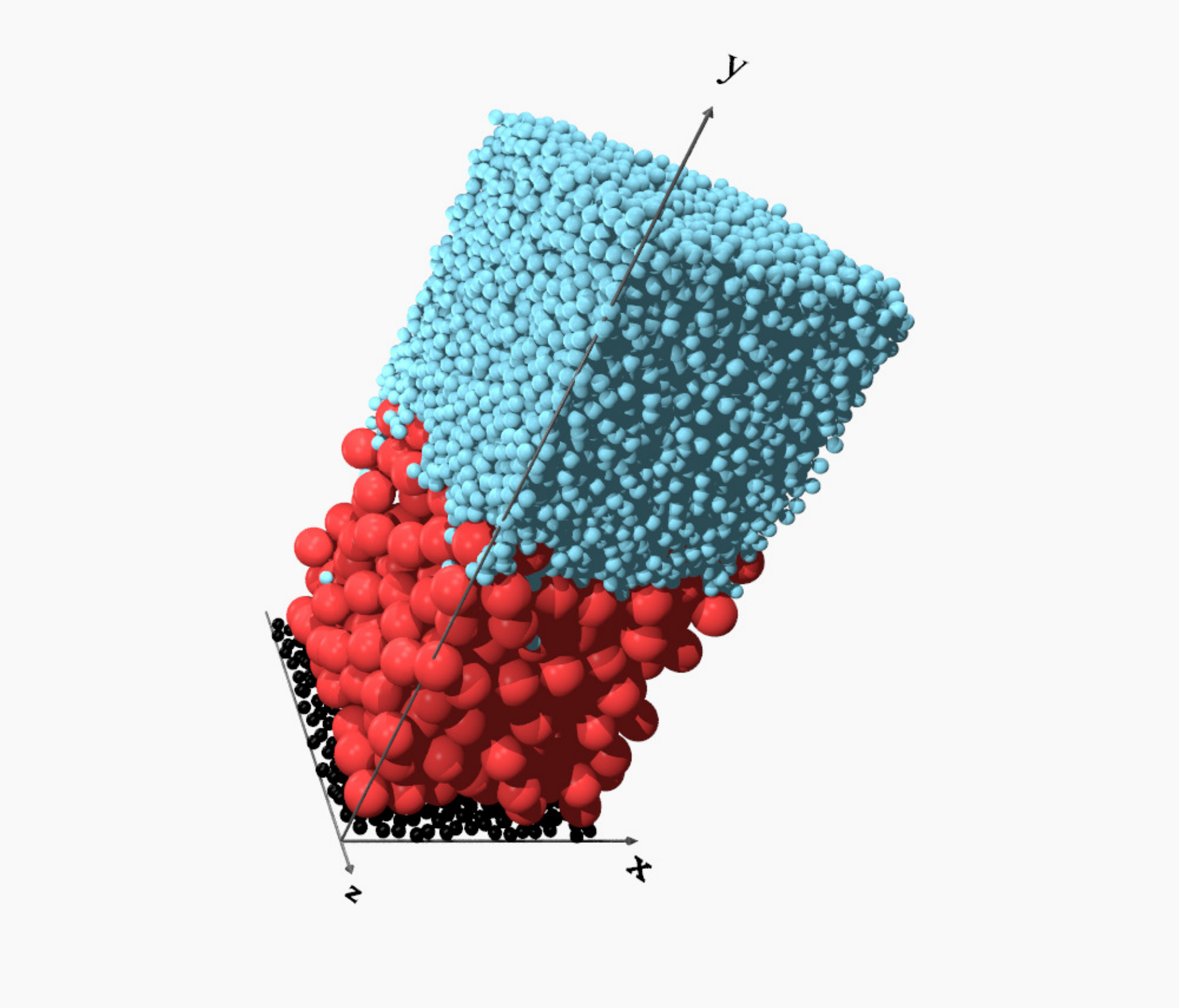}\put(-100,160){(d)}
    \caption{DEM snapshots of equal volume binary mixtures flowing down a plane inclined at $\theta = 25^\circ$ started from a large near base initial state for different particle size ratios of (a) $r = 1.5$, (b) $r = 2.0$, (c) $r = 2.5$, and (d) $r = 3.0$. Red particles indicate larger particles, while blue particles represent smaller ones. The black particles are static particles that are used to form a rough bumpy base to minimize slip at the chute base.}
    \label{fig:DEM_snap_size_ratios}
\end{figure}

\newpage
\section{Instantaneous flow properties for $r = 1.5$ at $\theta = 25^\circ$}
\begin{figure}[h]
    \centering
    \includegraphics[scale=0.34]{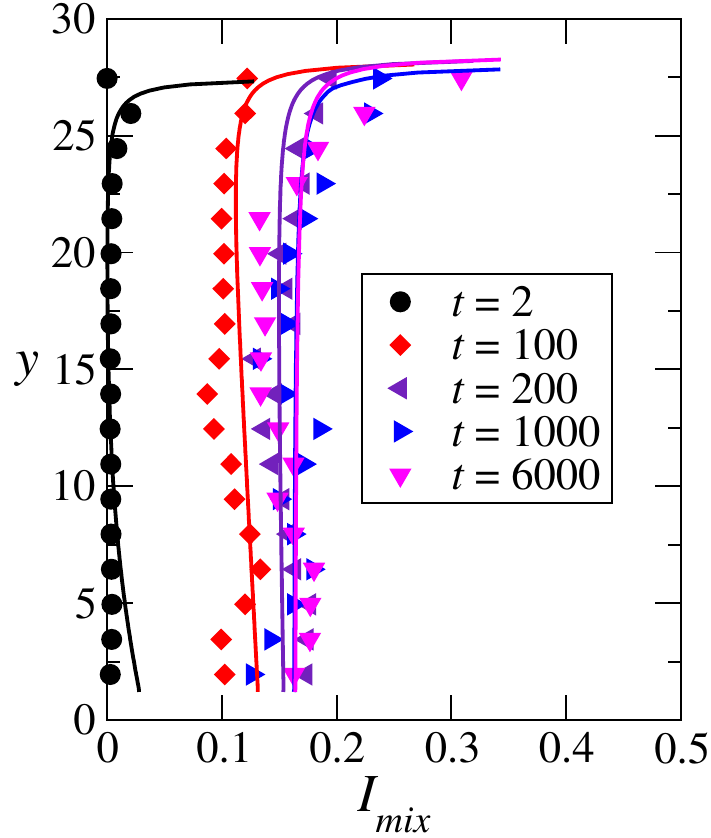}
    \put(-125,135){(a)}\quad \quad 
    \includegraphics[scale=0.34]{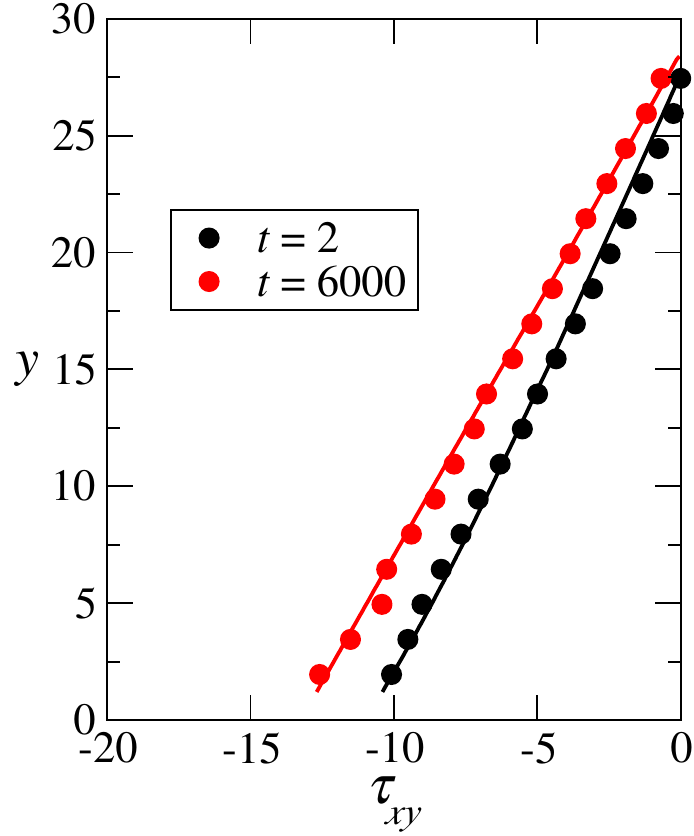}\put(-125,135){(b)}\hfill
    \caption{ Instantaneous profiles for (a) inertial number, and (b) shear stress for binary mixture of $50\% - 50\%$ large and small particles having size ratio $r = 1.5$ flowing over inclination angle $\theta = 25^o$. Symbols represent the DEM data while solid lines represents the continuum model predictions. \textcolor{black}{Shear stress is reported in dimensionless form using $m g/d^2$ as the reference scale, where $m$ and $d$ denote the mass and diameter of small particles, respectively.} }  
\label{fig:r_1.5_theta_25_instant_flowProp}
\end{figure}

\begin{figure}[h]
    \centering
    \includegraphics[scale=0.40]{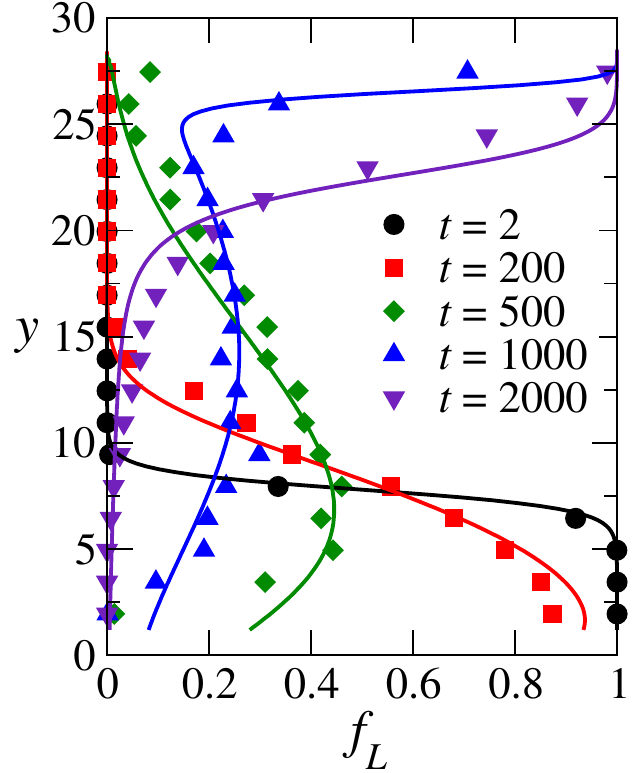}
    \put(-130,150){(a)}\quad \quad \quad \quad 
    \includegraphics[scale=0.40]{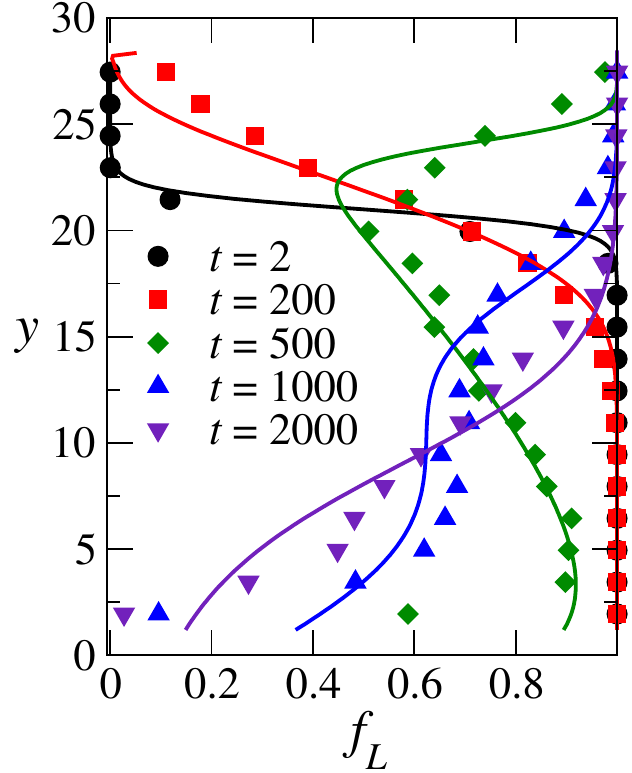}\put(-130,150){(b)}\hfill
    \caption{ Instantaneous concentration profiles for the large near base initial configuration of binary mixtures with (a) $f^T_L = 0.25$, and (b) $f^T_L = 0.75$, having size ratio $r = 1.5$ flowing over inclination angle $\theta = 25^o$. Symbols represent the DEM data and the solid lines represent the continuum model predictions.}  
    \label{fig:r_1.5_theta_25_instantfl_25_75}
\end{figure}

\clearpage
\section{Effect of size ratio on segregation}

\begin{figure}[h]
    \centering
     \includegraphics[scale=0.32]{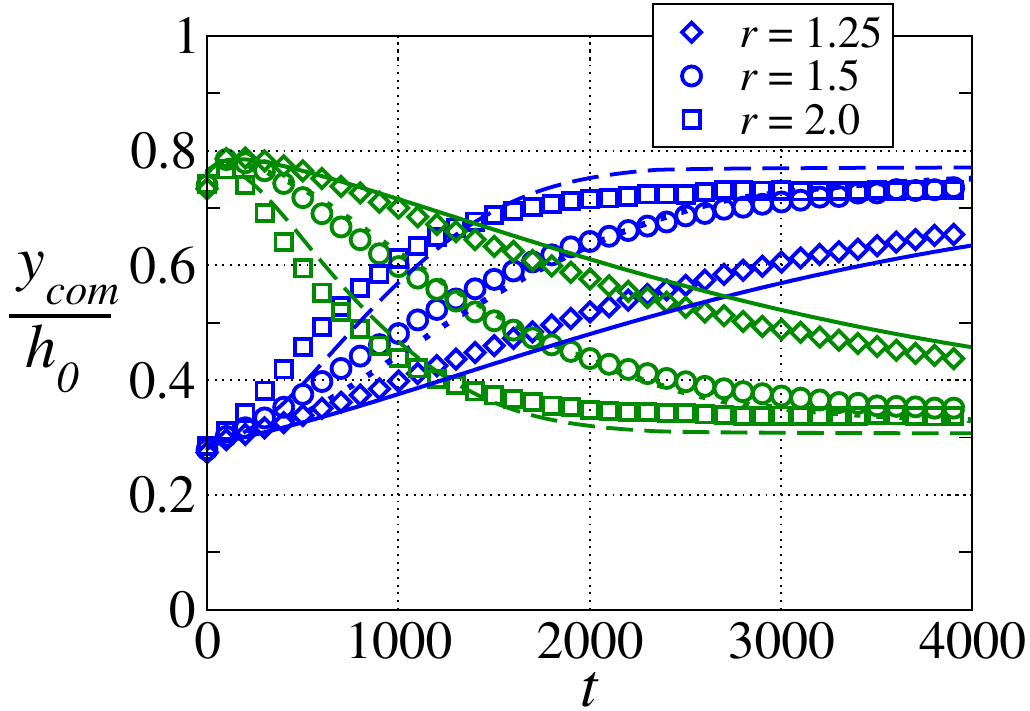}\put(-155,110) {(a)}\quad 
       \includegraphics[scale=0.32]{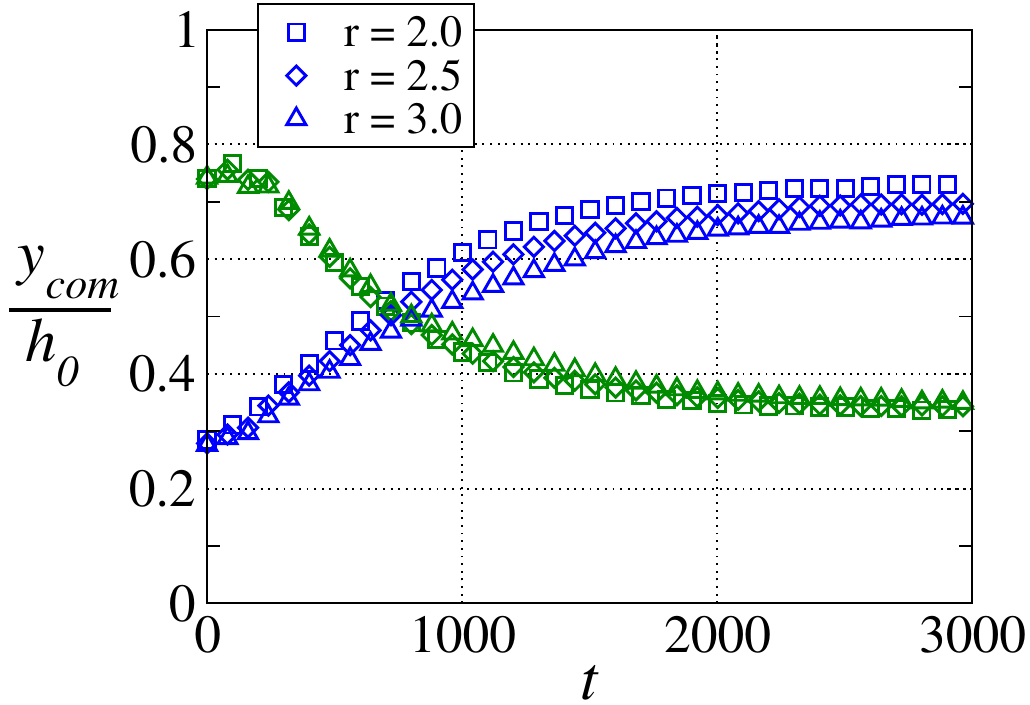}\put(-155,110) {(b)}\quad
         \includegraphics[scale=0.32]{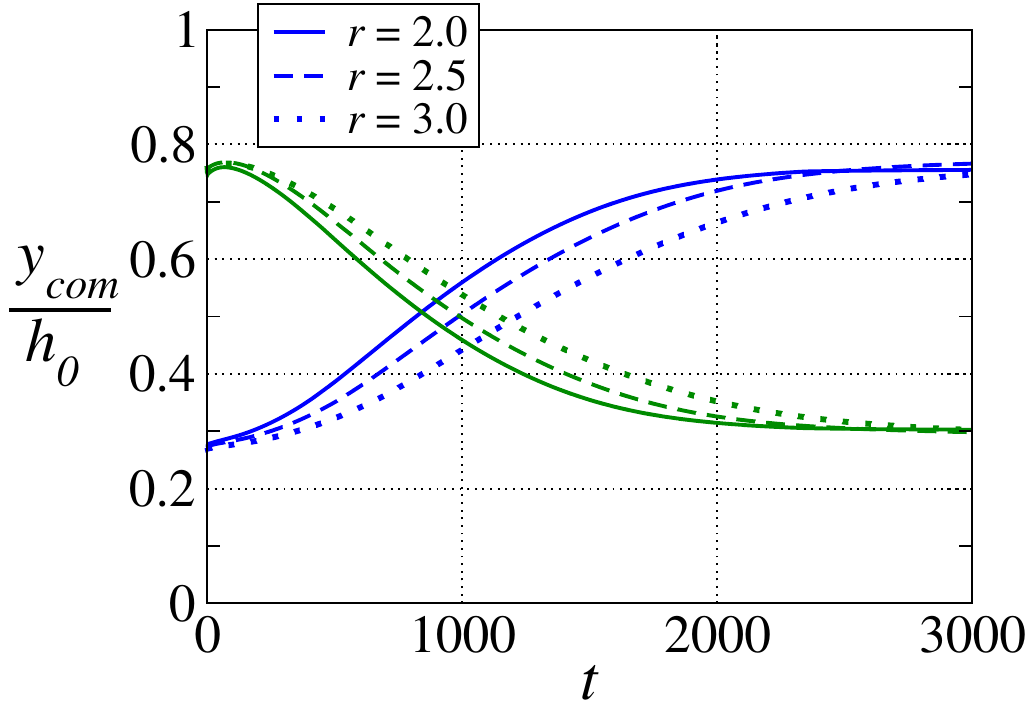}\put(-155,110) {(c)}
    \caption{(a) Comparison of time evolution of species center of mass position $y_{com}$ scaled by the initial layer height $h_0$ for size ratios (a) $r = 1.25$, $1.5$ and $2.0$, (b~$\&$~c) $r = 2.0$, $2.5$ and $3.0$. Symbols represent the DEM data and lines represent the continuum model predictions.}
    \label{fig:size_effect_r_1.25_1.5_2_2.5_3}
\end{figure}

\section{Results for different mixture compositions for large size ratios}
\begin{figure}[h]
    \centering
    \includegraphics[scale=0.35]{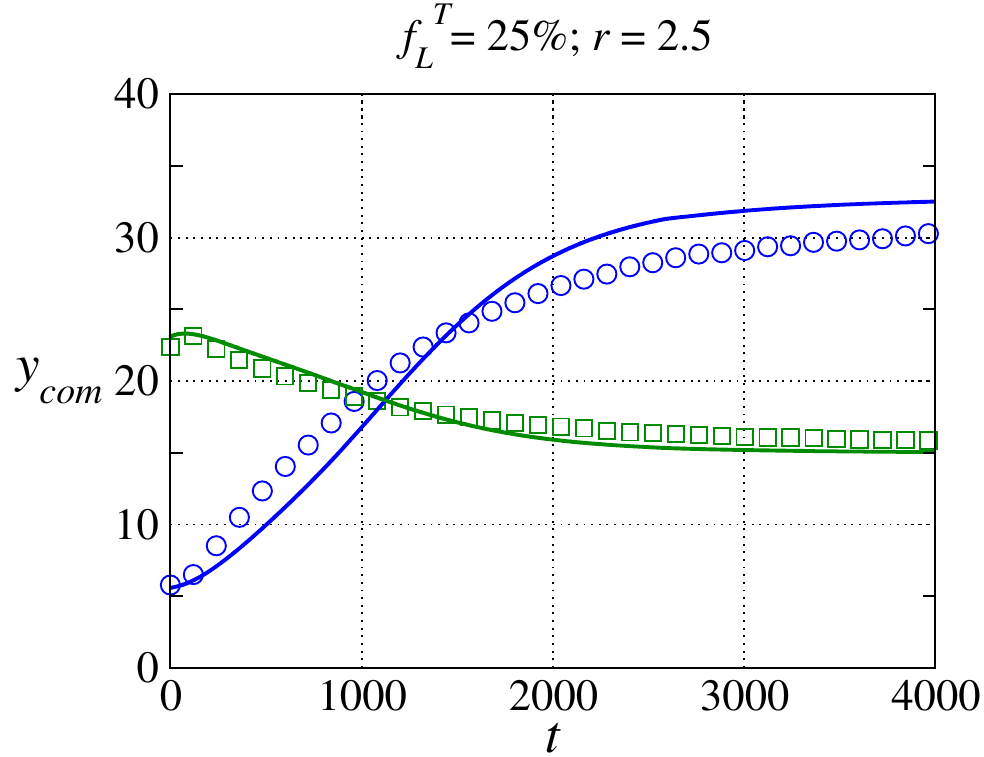}\put(-160,110){(a)} \quad 
      \includegraphics[scale=0.35]{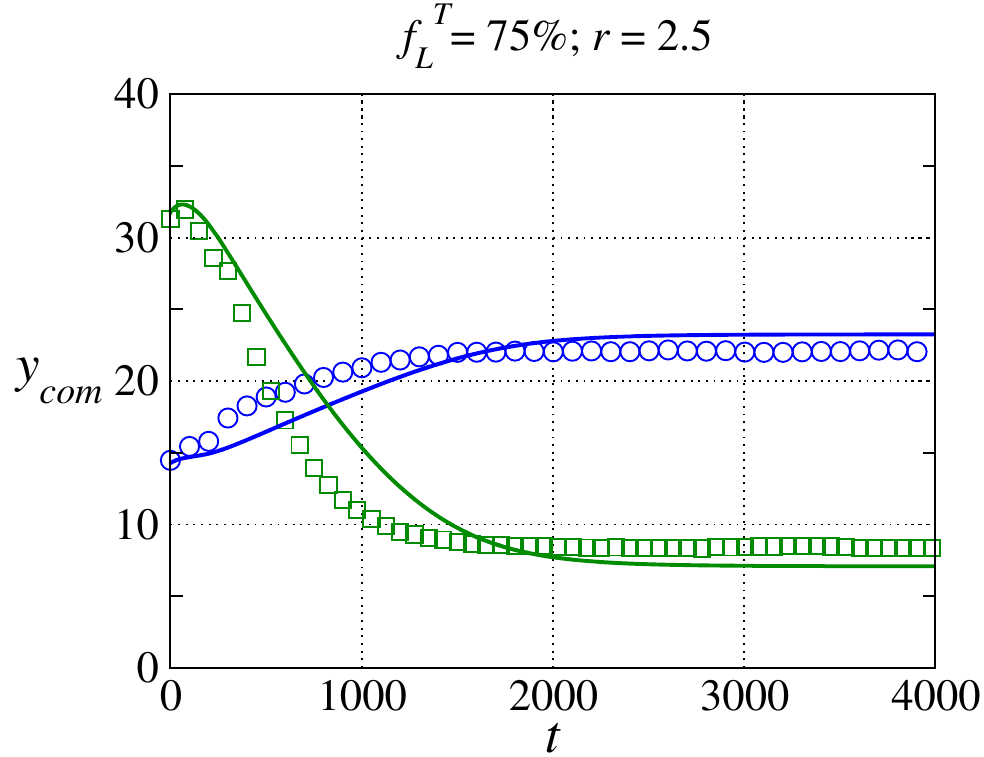}\put(-160,110){(b)} \\
 \includegraphics[scale=0.35]{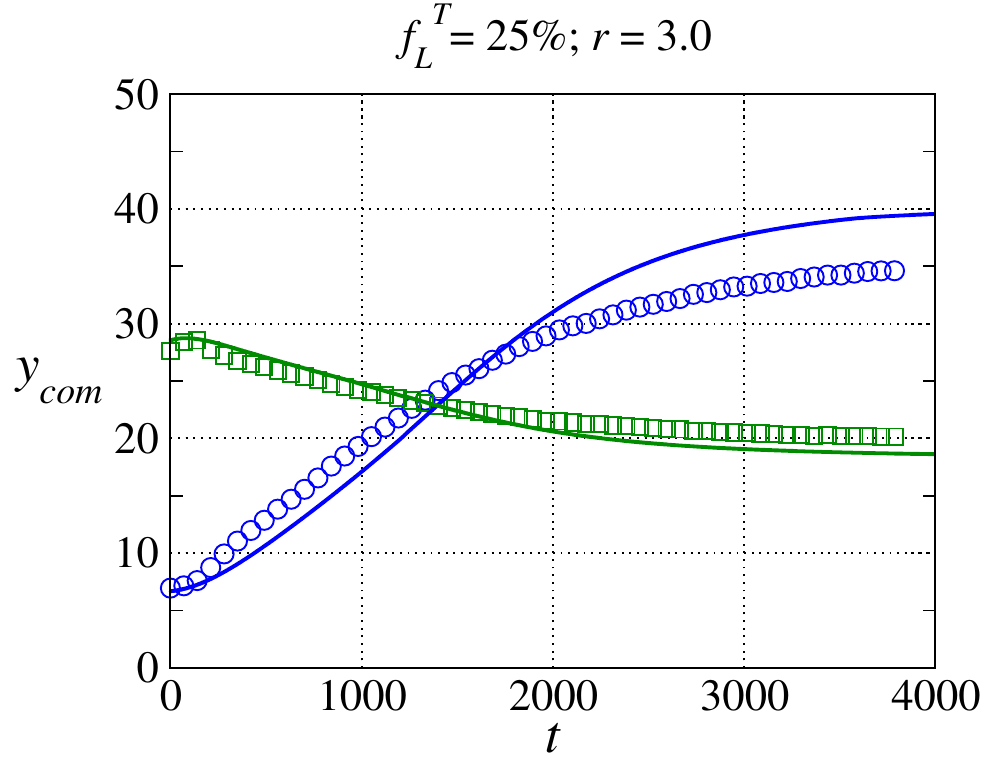}\put(-160,110){(c)} \quad
      \includegraphics[scale=0.35]{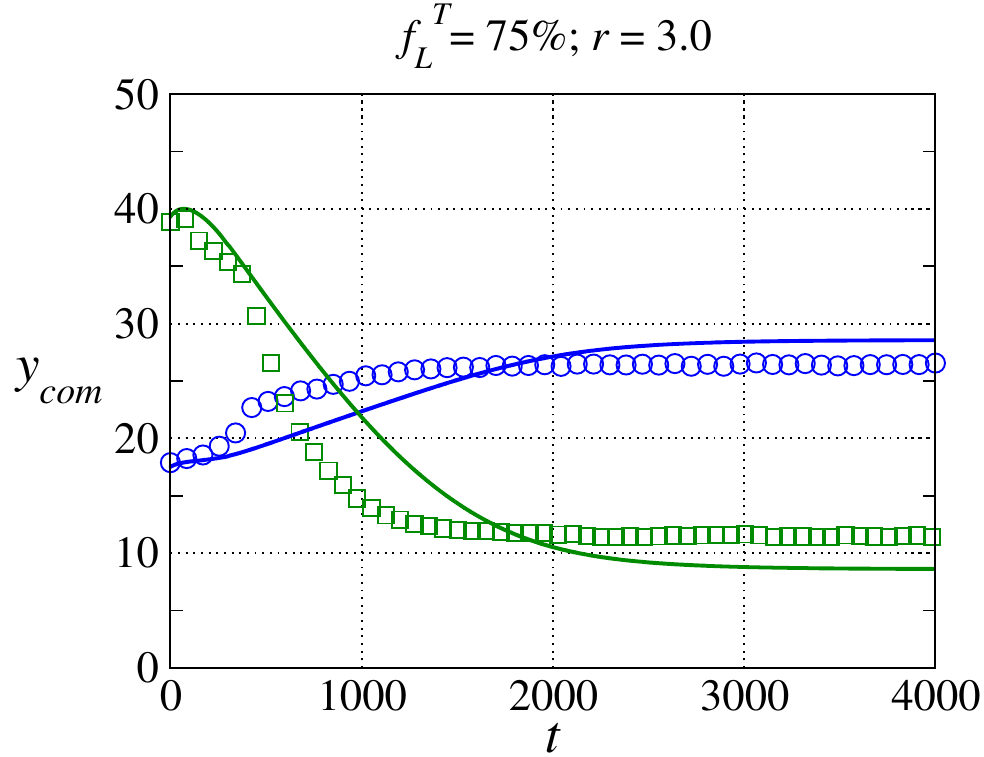}\put(-160,110){(d)}
    \caption{Variation of centre of mass of both large and small species of size ratio $r = 2.5$ for different mixture compositions, (a) $f^T_L = 0.25$, and (b) $f^T_L = 0.75$. (c) and (d) report the same data for $r = 3.0$.}
    \label{fig:flt_0.25_0.75_r_2.5_3}
\end{figure}

\end{document}